\documentclass[apj]{emulateapj}

\usepackage{wasysym}
\usepackage{graphicx}

\usepackage[normalem]{ulem}
\usepackage{color}
\definecolor{red}{rgb}{1.0,0.0,0.0}

\shorttitle{Resolved Debris Disk of HD 111520}
\shortauthors{Draper et al.}

\begin{document}

\title{The Peculiar Debris Disk of HD 111520 as Resolved by the Gemini Planet Imager}

\author{Zachary H. Draper\altaffilmark{1,2}, Gaspard Duch\^{e}ne\altaffilmark{3,4}, Maxwell A. Millar$-$Blanchaer\altaffilmark{5,6}, Brenda~C.~Matthews\altaffilmark{2,1}, Jason~J.~Wang\altaffilmark{3}, Paul~Kalas\altaffilmark{3}, James~R.~Graham\altaffilmark{3}, Deborah~Padgett\altaffilmark{7}, S.~Mark Ammons\altaffilmark{8}, Joanna Bulger\altaffilmark{9}, Christine~Chen\altaffilmark{10}, Jeffrey~K.~Chilcote\altaffilmark{6}, Ren\'{e}~Doyon\altaffilmark{11}, Michael~P.~Fitzgerald\altaffilmark{12}, Kate~B.~Follette\altaffilmark{13}, Benjamin~Gerard\altaffilmark{1,2}, Alexandra~Z.~Greenbaum\altaffilmark{14,10}, Pascale~Hibon\altaffilmark{15}, Sasha~Hinkley\altaffilmark{16}, Bruce~Macintosh\altaffilmark{13}, Patrick~Ingraham\altaffilmark{17} David~Lafreni\`{e}re\altaffilmark{11}, Franck~Marchis\altaffilmark{18}, Christian~Marois\altaffilmark{2,1}, Eric~L.~Nielsen\altaffilmark{18,13}, Rebecca~Oppenheimer\altaffilmark{19}, Rahul~Patel\altaffilmark{20}, Jenny~Patience\altaffilmark{21}, Marshall~Perrin\altaffilmark{10}, Laurent Pueyo\altaffilmark{10}, Abhijith~Rajan\altaffilmark{21}, Julian~Rameau\altaffilmark{11},  Anand~Sivaramakrishnan\altaffilmark{10}, David~Vega\altaffilmark{18}, Kimberly~Ward-Duong\altaffilmark{21} and Schuyler~G.~Wolff\altaffilmark{14,10}}{
\affil{$^{1}$Department of Physics and Astronomy, University of Victoria, 3800 Finnerty Rd, Victoria, BC V8P 5C2, Canada}
\affil{$^{2}$Herzberg Astronomy \& Astrophysics, National Research Council of Canada, 5071 West Saanich Road., Victoria, BC V9E 2E7, Canada}
\affil{$^{3}$Department of Astronomy, UC Berkeley, Berkeley CA, 94720, USA}
\affil{$^{4}$Universit\'{e} Grenoble Alpes / CNRS, Institut de Plan\'{e}tologie et d'Astrophysique de Grenoble, 38000 Grenoble, France}
\affil{$^{5}$Department of Astronomy \& Astrophysics, University of Toronto, Toronto ON M5S 3H4, Canada}
\affil{$^{6}$Dunlap Institute for Astronomy \& Astrophysics, University of Toronto, 50 St. George St, Toronto ON M5S 3H4, Canada}
\affil{$^{7}$NASA Goddard Space Flight Center, 8800 Greenbelt Road, Greenbelt, MD 20771, USA}
\affil{$^{8}$Lawrence Livermore National Lab, 7000 East Ave., Livermore, CA 94551, USA}
\affil{$^{9}$Subaru Telescope, NAOJ, 650 North A’ohoku Place, Hilo, HI 96720, USA}
\affil{$^{10}$Space Telescope Science Institute, 3700 San Martin Drive, Baltimore MD 21218 USA}
\affil{$^{11}$Institut de Recherche sur les Exoplan\`{e}tes, D\'{e}partment de Physique, Universit\'{e} de Montr\'{e}al, Montr\'{e}al QC H3C 3J7, Canada}
\affil{$^{12}$Department of Physics and Astronomy, UCLA, Los Angeles, CA 90095, USA}
\affil{$^{13}$Kavli Institute for Particle Astrophysics and Cosmology, Stanford University, Stanford, CA 94305, USA}
\affil{$^{14}$Physics and Astronomy Department, Johns Hopkins University, Baltimore MD, 21218, USA}
\affil{$^{15}$European Southern Observatory, Casilla 19001-Santiago 19-Chile}
\affil{$^{16}$University of Exeter, Astrophysics Group, Physics Building, Stocker Road, Exeter, EX4 4QL, UK}
\affil{$^{17}$Large Synoptic Survey Telescope, 950 N Cherry Av, Tucson AZ 85719, USA}
\affil{$^{18}$SETI Institute, Carl Sagan Center, 189 Bernardo Avenue, Mountain View, CA 94043, USA}
\affil{$^{19}$American Museum of Natural History, New York, NY 10024, USA}
\affil{$^{20}$California Institute of Technology, Infrared Processing and Analysis Center, 770 South Wilson Avenue, Pasadena, CA, 91125}
\affil{$^{21}$School of Earth and Space Exploration, Arizona State University, PO Box 871404, Tempe, AZ 85287, USA}

\begin{abstract}
Using the Gemini Planet Imager (GPI), we have resolved the circumstellar debris disk around HD~111520 at a projected range of $\sim$30-100 AU in both total and polarized $H$-band intensity.  The disk is seen edge-on at a position angle of $~$165\degr\ along the spine of emission. A slight inclination or asymmetric warping are covariant and alters the interpretation of the observed disk emission. We employ 3 point spread function (PSF) subtraction methods to reduce the stellar glare and instrumental artifacts to confirm that there is a roughly 2:1 brightness asymmetry between the NW and SE extension. This specific feature makes HD~111520 the most extreme examples of asymmetric debris disks observed in scattered light among similar highly inclined systems, such as HD~15115 and HD~106906. We further identify a tentative localized brightness enhancement and scale height enhancement associated with the disk at $\sim$40 AU away from the star on the SE extension. We also find that the fractional polarization rises from 10 to 40\% from 0\farcs5 to 0\farcs8\ from the star. The combination of large brightness asymmetry and symmetric polarization fraction leads us to believe that an azimuthal dust density variation is causing the observed asymmetry. 
\end{abstract}


\section{Introduction}

Improved resolution in debris disk imaging has made it possible to uncover many instances of complex morphologies which deviate from the nominally pervasive symmetric ring structures.  This offers important insights into the dynamical evolution of the planetary systems, since gaps and asymmetries will result from planet scattering, stellar fly-bys, and ISM interactions (for a review see \citealt{BM14a}{}). Investigations into these important case studies can determine how planetary architectures shape debris disks, or even create them, through planetary stirring of planetesimals \citep{AM09}. Even when the planets themselves may be unseen, important constraints can be made based on the disks' structure \citep{SE12}. 

\begin{figure*}
\centering
\includegraphics[trim={0mm 0mm 8cm 0mm},clip,height=0.23\textheight]{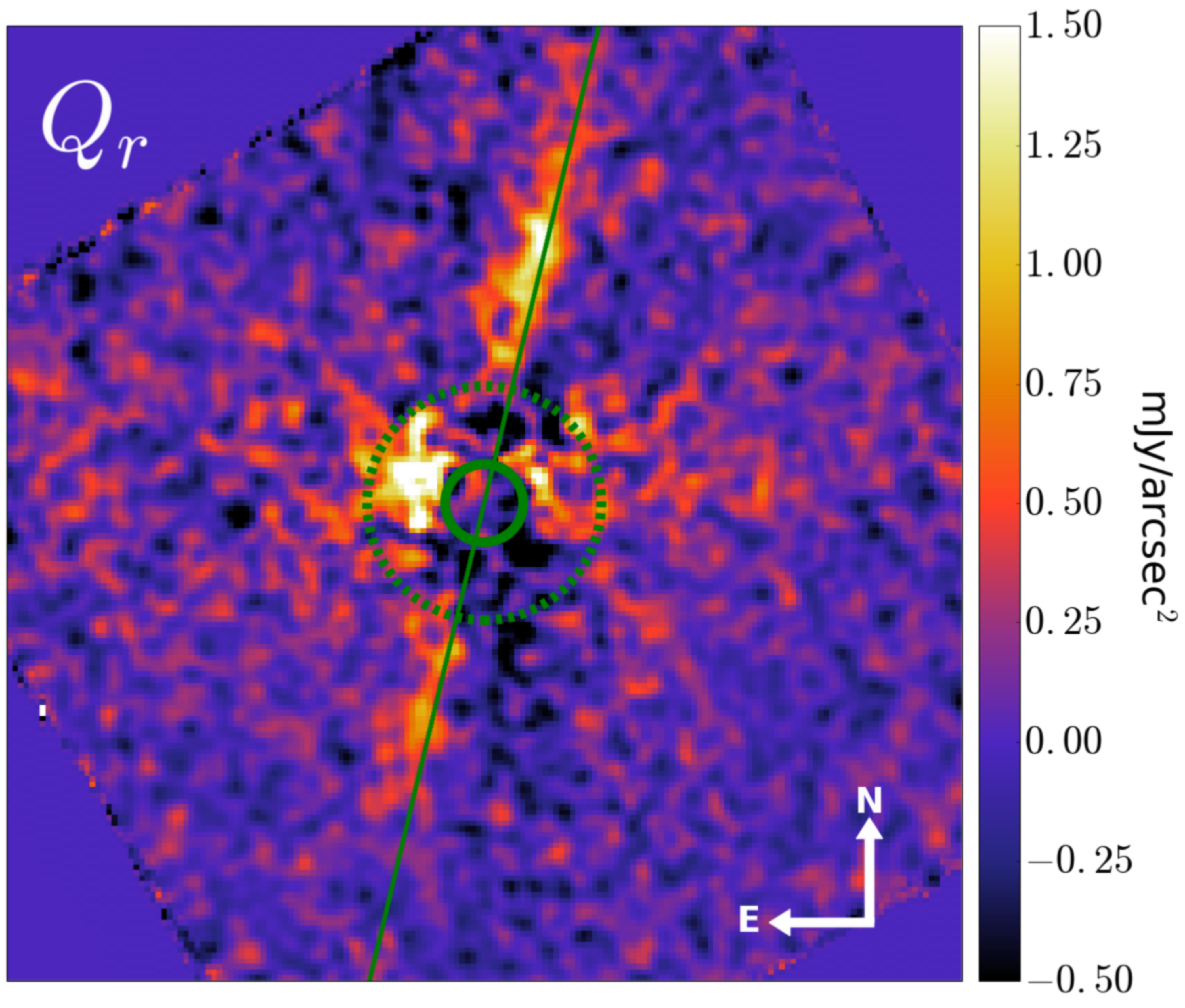}
\includegraphics[height=0.23\textheight]{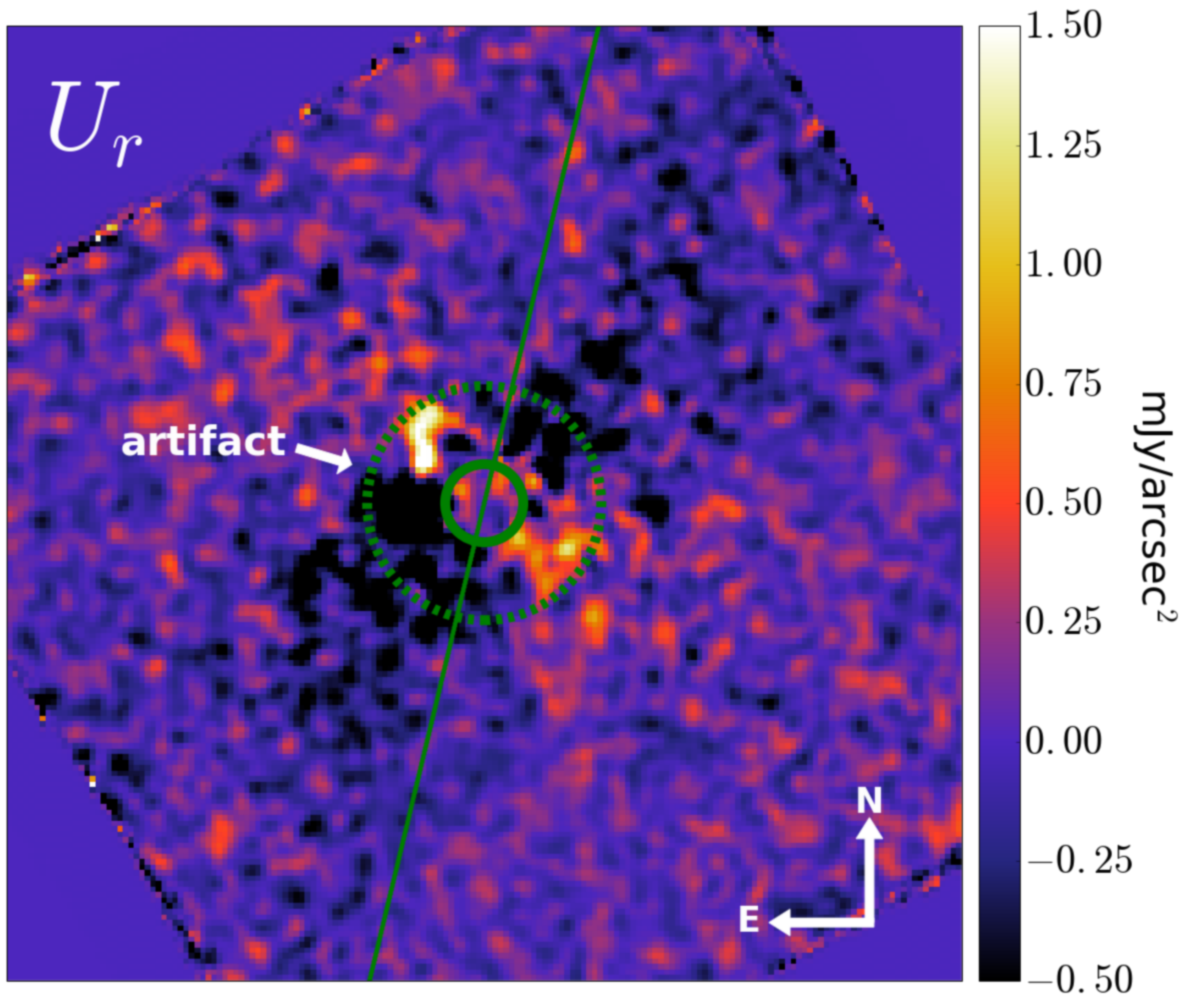}
\includegraphics[height=0.23\textheight]{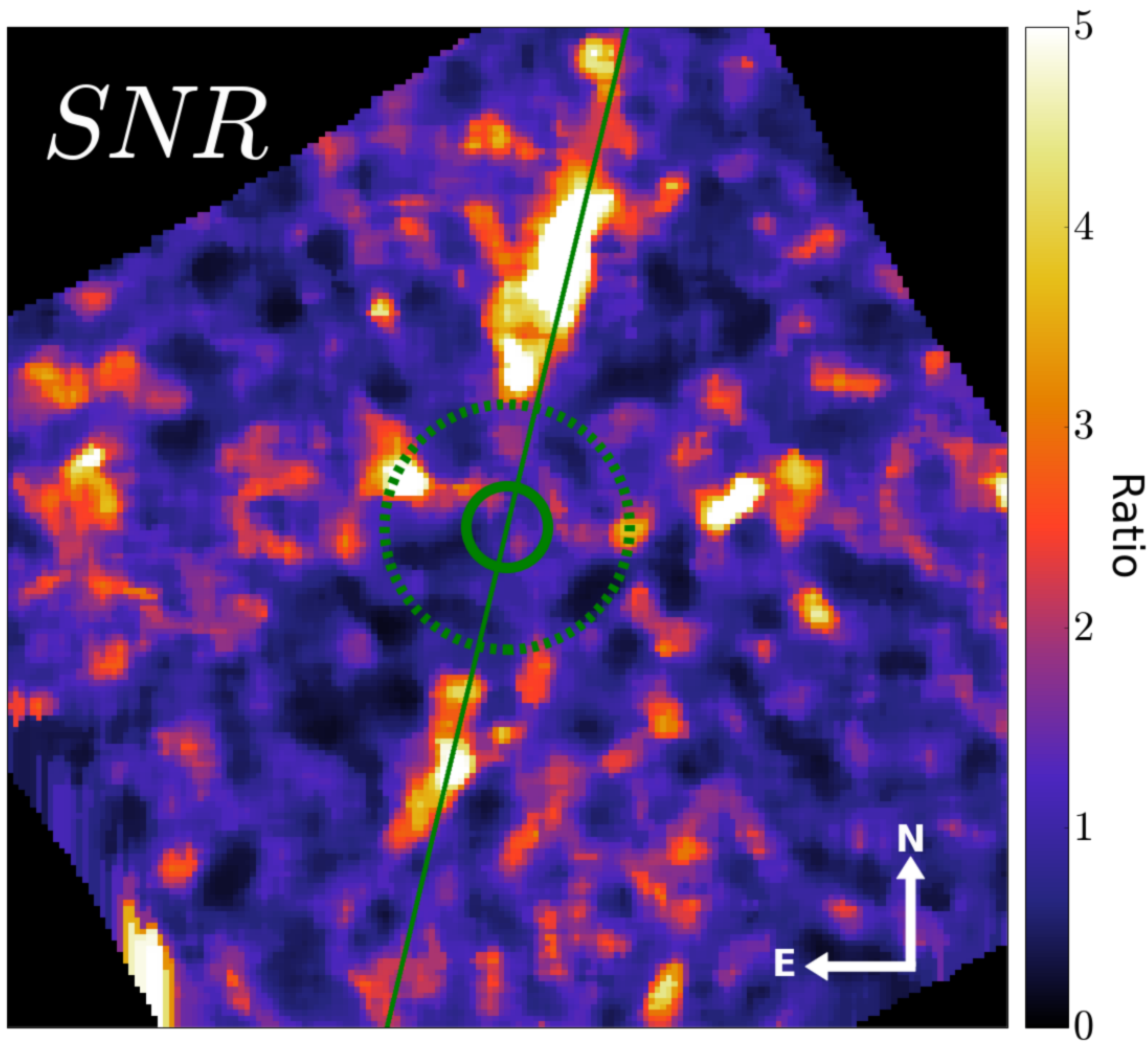}
 \caption{$H$-band radial Stokes polarized intensity. The coronagraph is marked by a solid green circle. The FOV of the images is cropped to $2.4\arcsec\times2.4\arcsec$. The dashed green circle denotes a region with enhanced noise out to 0.3$\arcsec$ in radius from the center. (Left) Radial Stokes $Q_r$ showing that the disk emission is aligned along a position angle (PA) of 165\degr\ centered at the star, illustrated by the green line. (Center) Radial Stokes $U_r$ shows polarized light from non-astrophysical sources (assuming single scattering) and is therefore an estimate of the noise in the data. Both $Q_r$ and $U_r$ images are shown using the same color scale.
 A large artifact $\sim$0\farcs1-0\farcs3 to the East of the coronagraph appears in both $Q_{r}$ and $U_{r}$ and is therefore likely an instrumental effect. (Right) An SNR map showing the detection of the disk.}
\label{fig:pol}
\end{figure*}

This paper presents resolved imaging from GPI and evidence for strong asymmetry in the disk around HD~111520 (HIP~62657) which is seen from 0\farcs3--1\farcs0. GPI is an instrument designed to detect scattered light from dust grains and emission from exoplanets in the near-IR at close separations around nearby stars \citep{BM14}. HD~111520 is an F5V star and has been identified as a member of the Lower Centaurus Crux (LCC) in the Scorpius-Centaurus Association through Hipparcos proper motions \citep{zee99}. Stellar parameter estimates have ranged from $6500-6750$ K surface temperature, $2.6-2.9$ $L_{\astrosun}$, and $1.3-1.4$ $M_{\astrosun}$\citep{CC14,pec12,mss78}. The distance to the system was measured to be $108\pm12$~pc \citep{lee07}, which we adopt throughout this study.  The median age of the LCC for F-type stars is 17$\pm$5~Myr \citep{pec12}. 

An IR-excess was first associated with the star by \cite{chen11} based on \textit{Spitzer} MIPS data which derived a dust radius of 48 AU from a fit to the the effective temperature of a single blackbody.  In combination with \textit{Spitzer} IRS, multiple temperature components have been fit with grain emissivity models to give an inner disk of 115~K at a radius of 16.3~AU and an outer disk of 51~K at 212~AU \citep{CC14}.  Subsequent detailed grain model fits have been done to IRS spectra to give estimates of an inner disk at 1~AU and an outer disk of 20~AU \citep{TM15}, although this model greatly underpredicts the 70$\mu$m flux, requiring another outer component. These discrepancies in SED fitting are primarily due to model degeneracies in the absence of a resolved image of the disk structure.  The disk around HD~111520 was first resolved in optical scattered light by HST to have a large 5:1 brightness asymmetry with emission extending from $\sim$1\arcsec-5\arcsec (or $\sim$110-550 AU) from the star \citep{DP15}.   Indeed, all SED models predict a dust location that is well inside the inner working angle of the discovery HST images, but within the GPI field-of-view (FOV), underlining the importance of GPI for understanding warm debris disk dust. We therefore present GPI data which resolves the disk inside 1\arcsec\ to better probe the structure of the disk.

\begin{figure} 
\centering
\includegraphics[height=0.29\textheight]{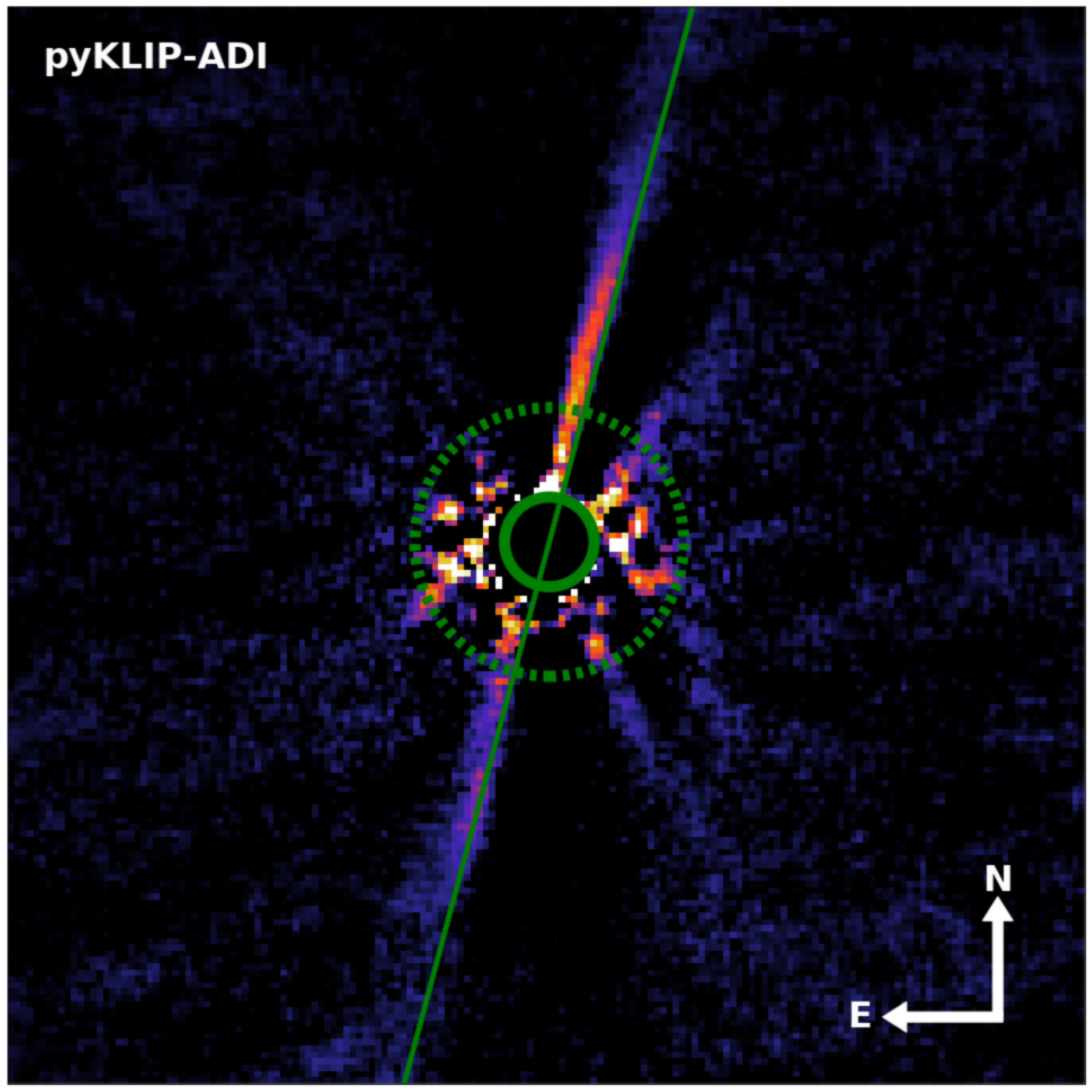}
\includegraphics[height=0.29\textheight]{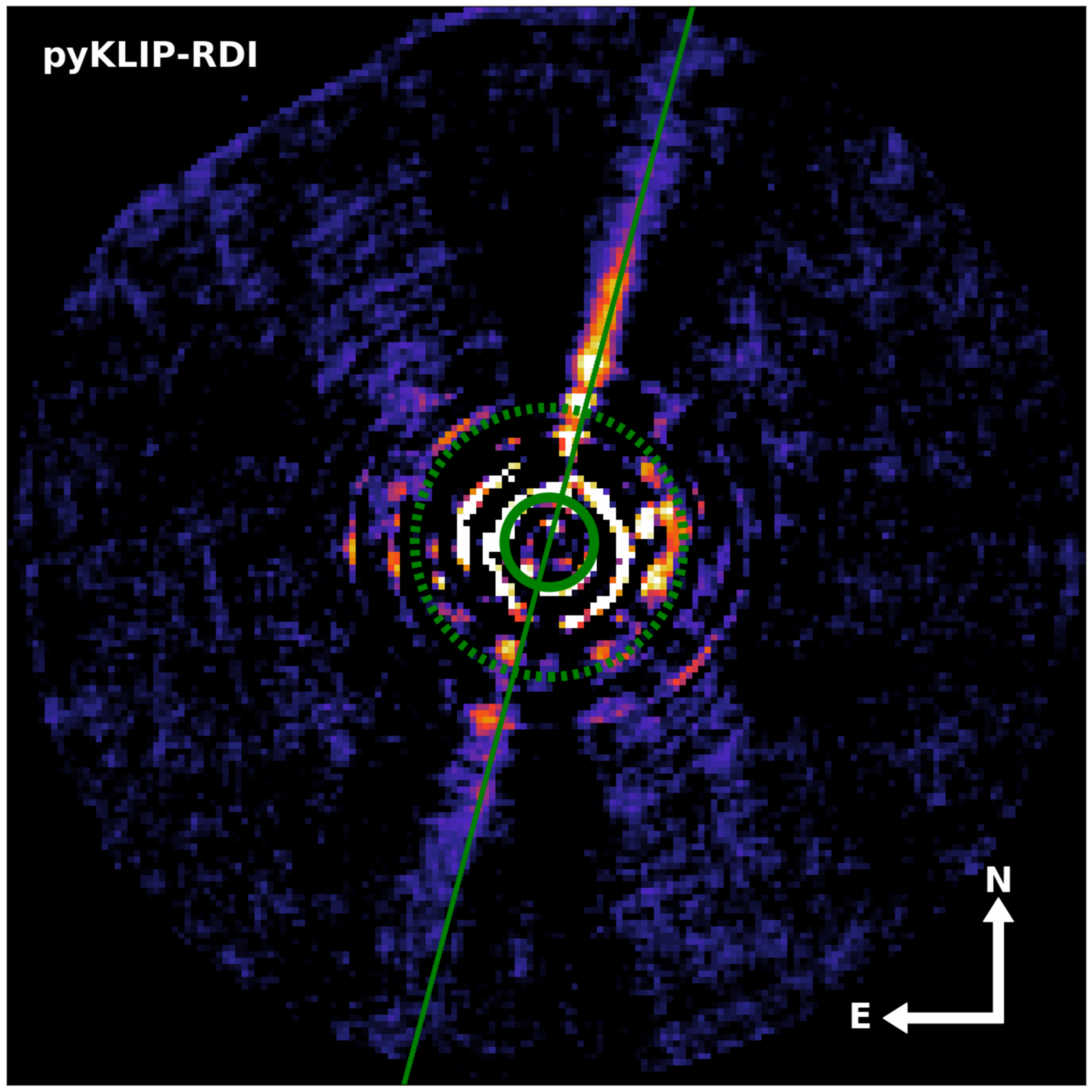}
\includegraphics[height=0.29\textheight]{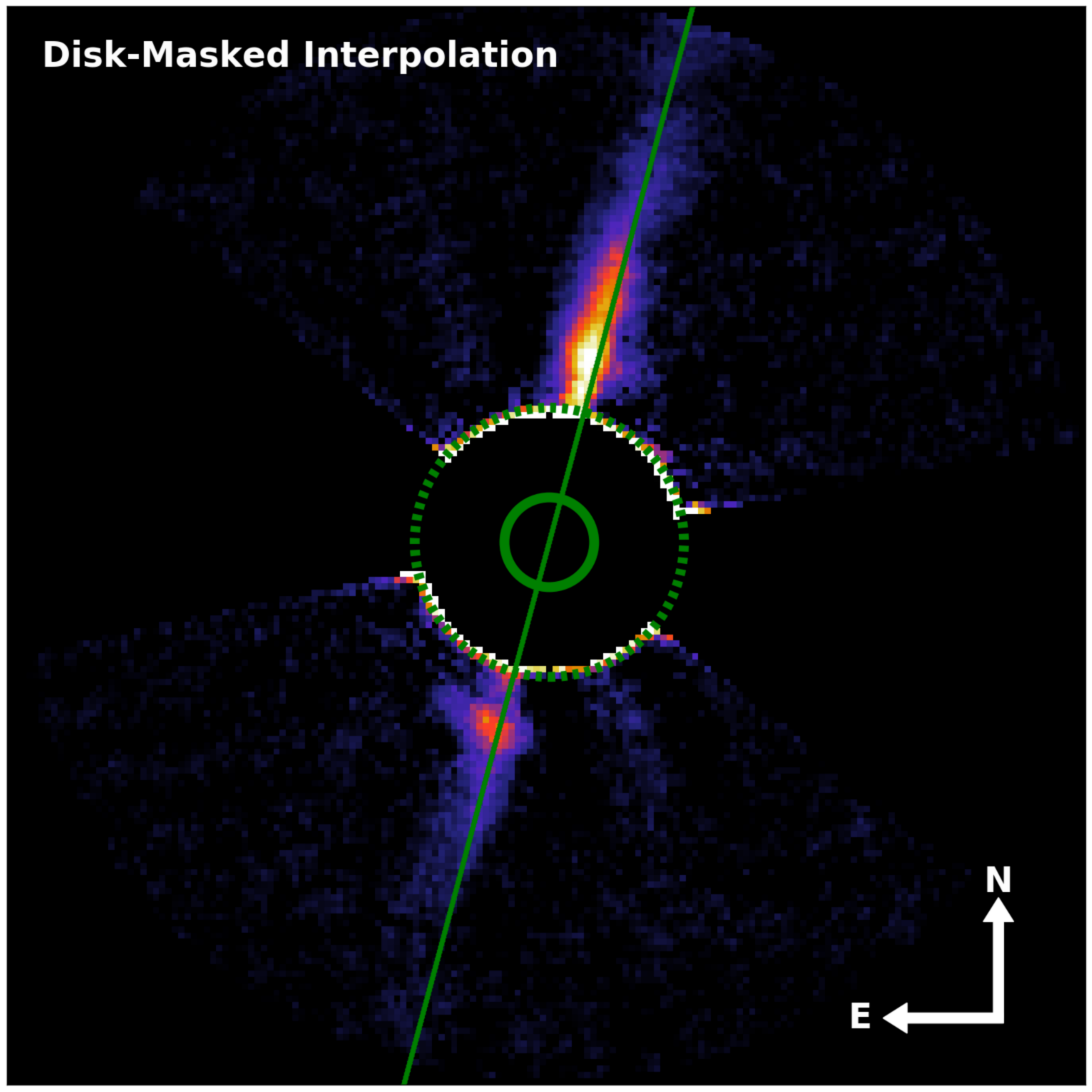}

\caption{Collapsed $H$-band spectral mode data reduced using various PSF subtraction methods. The FOV of the images is cropped to 2\farcs4$\times$2\farcs4. (Top) Reduced with an ADI-only reduction with pyKLIP. (Center) PSF-subtracted data using a PSF library from GPI Exoplanet Survey data as a reference for a pyKLIP reduction. (Bottom) PSF subtracted by interpolating over disk-masked data as done in \cite{per14}. The solid green circle denotes 0\farcs1 which is obstructed by the coronagraph.  The dashed green circle denotes a 0\farcs3 radius inside of which large artifacts are present in the all of the PSF reductions.  The solid line denotes the primary plane of the disk major axis of the emission along a PA of 165\degr.}

\label{fig:psf}
\end{figure}

\section{Observations and Data Reduction}
\label{data_reduc}

On the night of 2015-07-02, data were taken as part of the GPI Exoplanet Survey \citep[GPIES,][]{BM14}. Weather conditions were good with DIMM (Differntial Image Motion Monitor) seeing at $\sim$1\arcsec\ and MASS (Multi-Aperture Scintillation Sensor) seeing at $\sim$0\farcs5. A total of forty-one 60~s exposures were taken in $H$-band spectral mode (R$\sim$45) with a total of $\sim$35$^{\circ}$ of field rotation.  In addition, eleven 60~s exposures in $H$-band polarization mode were taken for a `snap-shot' observation amounting to $7^{\circ}$ of rotation. The field rotation allows for Angular Differential Imaging (ADI) to subtract the instrument PSF \citep{CM06}. The pixel scale of GPI data is $14.166\pm0.007$ milli-arcseconds on the sky \citep[updated from][]{QK14}. The data were reduced using primitives in the GPI Data Reduction Pipeline \citep[see][and references therein]{per14}. 

For polarimetry mode data, the light is split by a Wollaston prism into two orthogonal linear polarization states that are modulated by a rotating, achromatic half-wave plate. A typical observing sequence involves observations in sets of four different wave plate orientations, which are then combined to produce a Stokes datacube \citep{MP15}. First, the raw frames are dark subtracted and `destriped' using Fourier-filtered raw detector images to remove instrumental microphonic noise \citep{PI14}. The microlenslet spot locations from a calibration file are corrected for instrument flexure with a cross-correlation algorithm \citep{ZD14}. The raw data are then converted to a polarization datacube, where the third dimension contains the two orthogonal polarization states. Systematic variations in the polarization pairs and bad pixels are cleaned by a modified double difference algorithm \citep{per14}. A geometric distortion correction was also applied \citep{QK14}.  The data are then smoothed by a Gaussian kernel with a width equivalent to a nearly diffraction limited GPI PSF (FWHM = 3 pixels). By measuring the fractional polarization behind the occulted spot, the instrumental polarization is measured and subtracted off from each pixel based on its total intensity \citep{MB15}. Following \cite{LH15}, flux calibration was performed measuring the photometry of the satellite spots with elongated apertures with a known conversion to compare with the 2MASS magnitude for the star ($7.830\pm0.057$~mag or $0.756\pm0.040$~Jy; below 2MASS saturation limits; \citealt{CR03}). All of the polarization datacubes were then combined via a singular value decomposition method to create a Stokes datacube \citep{per14}.  Finally the Stokes cube was converted to the radial Stokes convention:  $[I,Q,U,V] \rightarrow [I,Q_r,U_r,V]$ \citep{HS06}. The star location, which is used as the origin of the transformation, is measured using a radon transform-based algorithm that takes advantage of the elongated satellite spots \citep{JW14,LP15}. The final $Q_r$ and $U_r$ images can be seen in Fig.~\ref{fig:pol}. 

For the spectroscopy mode data, the raw dispersed frames were dark subtracted, corrected for bad pixels, and `destriped' \citep{PI14}. A wavelength calibration using an Ar arc lamp  was taken just prior to the observations and corrected using a repeatable flexure model of the instrument as a function of telescope elevation \citep{SW14}. In this case, the correction amounted to a negligible change from the nominal wavelength calibration. To extract into a 3D spectral datacube, a box aperture method was used \citep{JM14s}. There were interpolation errors along the wavelength axis at the blue end of the data cubes, so the first three individual spectral channels (or 0.024 $\mu$m bandpass) were removed prior to collapsing the cube. A flat field image can have a pixel to pixel standard deviation on order of $\sim$10\% and therefore cannot explain surface brightness variations above this level. A microlens-PSF method \citep{PI14b,ZD14} was also used to optimize the flux extraction and reduce spaxel-to-spaxel noise (i.e. spectral pixels). These cubes did not have bad cube slices but yielded similar results for the PSF subtracted images. To remove persistent bad spaxels, they are identified as being discrepant from a spatial $3\times 3$ box median filtered image per wavelength slice and then smoothed by assigning it the median value of a $3\times 3 \times 3$ region within the cube centered on the bad spaxel. The satellite spots locations were identified after high pass filtering in order to derive the star location under the coronograph for each datacube.  The star centering accuracy is 0.05 pixels for satellite spots with SNR$>$20 for spectral datacubes \citep{JW14}. Our data has an SNR around 20 which can be up to 0.1 pixels or 1.4\,mas in astrometric precision. Finally, the data were flux calibrated using the satellite spots within the image and the target's 2MASS magnitude and spectral type (for bandpass color corrections) into surface brightness \citep{JW14}.  In all, varying the use of any of these data cube reduction steps did not significantly alter the resulting data cubes to level of spatial flux variation seen in \S \ref{radial}. Lenslet flat fielding tended to introduce more ``checkerboard'' or spaxel-to-spaxel noise in the data cubes, likely because they were obtained on a different night with a flexure shift causing the microlenslets to sample different pixels on the detector.  Therefore it was left out of the data reduction procedure. 

\section{PSF Subtraction}

The spectral mode cubes require PSF subtraction to remove instrumental scattered light and isolate the astrophysical emission. The spectral mode cubes were combined with pyKLIP \citep{JW15} using ADI-only mode of individual spectral channels \citep{CM06}. The resulting image from the collapsed cube is shown in the top panel of Fig.~\ref{fig:psf}. The KLIP algorithm uses a principal component analysis method, in concert with the angular rotation of the data sets, to determine the best PSF model to subtract \citep{RS12}.  A median of multiple iterations of pyKLIP using 41 KL mode basis vectors with annuli and angular subsections ranging from 5 to 18 equal subdivisions of the image, in both width and angular size, were combined to produce the final image.  

In order to confirm that the apparent NW to SE asymmetry seen in the pyKLIP reduction was not due to self-subtraction, we also applied a version of pyKLIP that used reference differential imaging (RDI). Instead of using the target dataset to construct the PSF, this method relied on an extensive broadband PSF library composed of observations of disk- and companion-free reference stars obtained during the GPIES campaign. Broadband images were created either by summing all the wavelength channels in spectroscopy mode datacubes or by summing the two orthogonal polarization states in polarimetry mode data. In this way, data from both observing modes can be used as broadband PSF references. At the time these reductions were carried out, the library consisted of approximately 7400 PSFs. All of the GPIES datacubes were reduced in a similar manner, following the standard reduction recipes (e.g. \S \ref{data_reduc}). For each spectroscopy mode datacube in the HD 111520 dataset, the 100 most correlated PSFs in the library were selected as reference PSFs and then processed using pyKLIP. The reduction used a combination of 3 and 6 pixel annuli and 10 KL modes, with vector lengths ranging from 7 to 49, which were averaged together to smooth out remaining artifacts. The result can be seen in the center panel of Fig.~\ref{fig:psf}. A more detailed description of the broadband PSF library will be discussed further in an upcoming paper (Millar-Blanchaer et al., in prep).

Another method we employed to preserve disk flux consists of subtracting a PSF model, interpolated from data which had the disk masked, before recombining the data set (bottom panel of Fig.~\ref{fig:psf}). The method is similar to the PSF subtraction technique used on GPI data of HR 4796A \citep{MP15}. Each spectral cube is summed along its wavelength axis to make a broadband image. A rectangular region encompassing the extent of the disk is masked. The PSF is sampled outside of the masked region to fit a low-order polynomial over the masked regions. The PSF model is smoothed with a median filter and subtracted from each image before recombining the data by derotating into the same frame of reference on the sky. Depending on the normalization, the absolute flux level can vary by $\sim$30\% but does not impart localized surface brightness variations, such that relative differences in surface brightness are preserved.

In general, a pyKLIP-ADI PSF subtraction performs best at subtracting the residual PSF but leads to many artifacts which are not ideal for extended sources (Fig. \ref{fig:psf}). Given the edge-on nature of the disk, disk self subtraction is present, but is not strong enough to preclude it from detection as it would be for a centrosymetric face-on disk. Also, ringing and radial spokes are noticeable artifacts of this type of PSF subtraction. The NW to SE brightness asymmetry persists when using fewer KL-modes but structure in the fainter SE is less apparent. Overall, this method leads to over subtraction especially on the faint SE extension (compare the different panels of Fig \ref{fig:psf}). Using a PSF library as references for the reduction greatly enhances the optimal subtraction but still leaves some of the KLIP artifacts. On the other hand, the masked PSF fitting leads to the least subtraction of the disk, though at the cost of a slightly larger inner working angle where residual artifacts dominate. We therefore use the latter method to measure the disk's surface brightness and morphology. Since the polarization mode dataset has fewer observations and less parallactic rotation, we use the spectral mode data to constrain the total intensity.  Through flux calibration between both modes respectively, we can compare the polarized intensity to the total intensity to get fractional polarization.

In the polarization mode with GPI, it is possible to isolate scattered light from a disk which is polarized and thereby remove the instrumental PSF which is assumed to be unpolarized. Light which scatters from optically thin dust around the star will have an electric field vector which is oriented centrosymmetrically around the star (parallel or orthogonal to rays emanating from the star), while residual polarized instrumental noise can be oriented at other orientations. In some cases optical depth effects, grain properties, and viewing geometry may impact this conclusion but it is robust for optically thin disks \citep{HC15}. As expected, the disk can be clearly seen in $Q_{r}$ with similar morphology to the total intensity (Fig. \ref{fig:pol} \& \ref{fig:psf}).  The $U_{r}$ image however shows correlated noise that we assume to be instrumental in origin just east of the coronagraph between 0\farcs1--0\farcs3.  The disk itself seen in $Q_{r}$ stands out above the noise shown in $U_{r}$ in relative strength and location.

\begin{figure}
\centering
\includegraphics[width=0.5\textwidth]{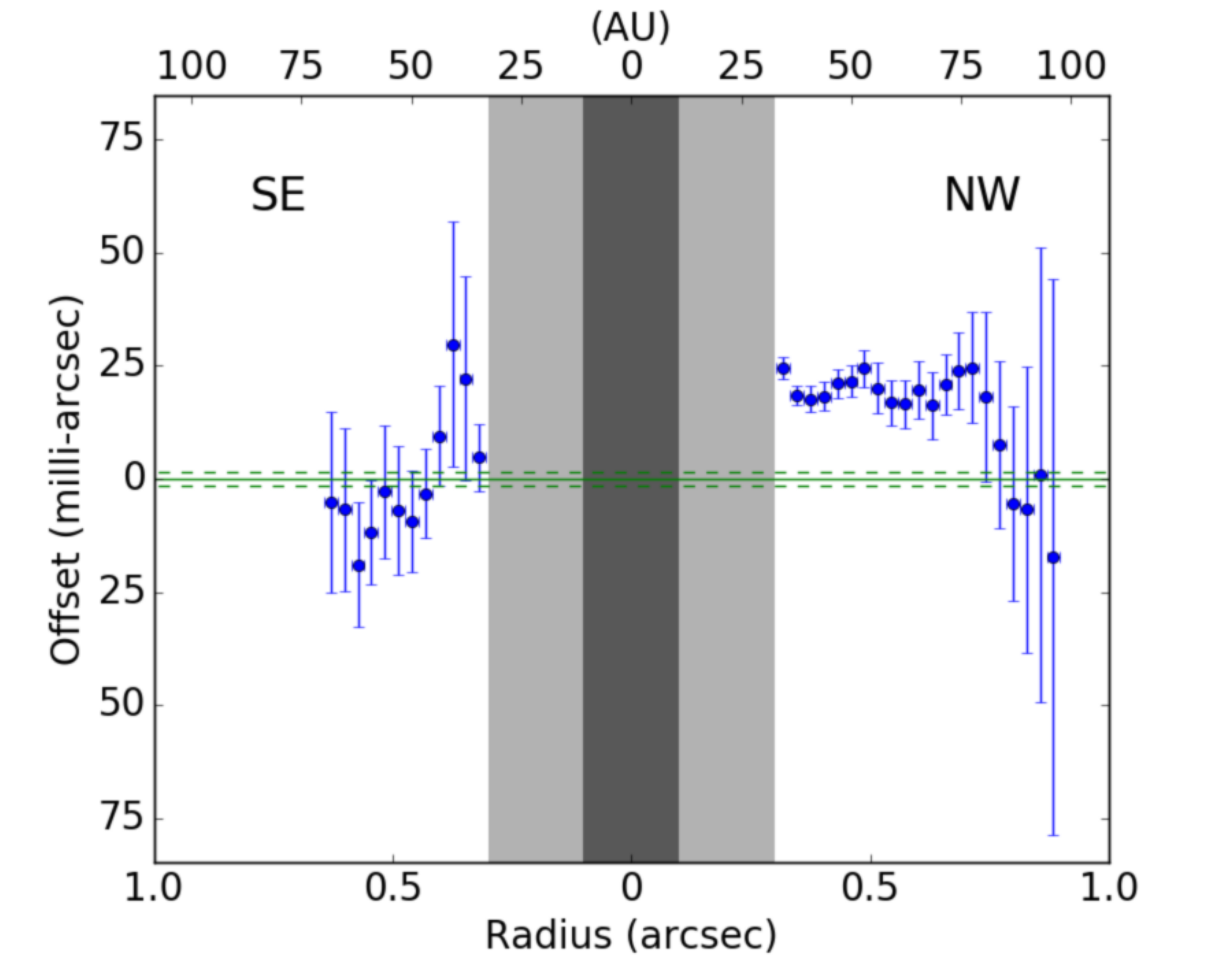}
\caption{
Vertical location of the peak disk emission along the spine relative to a line PA of 165\degr\ centered on the star. Light grey region indicates the region dominated by noise within 0\farcs3 and the dark grey shows the region under the chronograph at 0\farcs1 (See Fig. \ref{fig:psf}). The dashed green lines represent an upper limit on the uncertainty in the stellar position of 1.4\,mas. The disk emission does not appear to be arced, as would be the case for a symmetric ring slightly inclined from exactly edge-on, given the precision of our measurements (indicated by the FWHM/SNR as error bars). The slight offset we observe may nonetheless result from a few degree inclination of a closed ring disk, if the ring radius is significantly larger than the GPI FOV. Alternatively, a small warp producing an 'S' shape may be present but, given the brightness asymmetry, would not be as apparent. A localized offset on the SE extension though is apparent just inside 50 AU corresponding to an enhanced surface brightness feature (see Fig \ref{fig:radial}).}
\label{fig:offset}
\end{figure}

\begin{figure}
\centering
\includegraphics[width=0.5\textwidth]{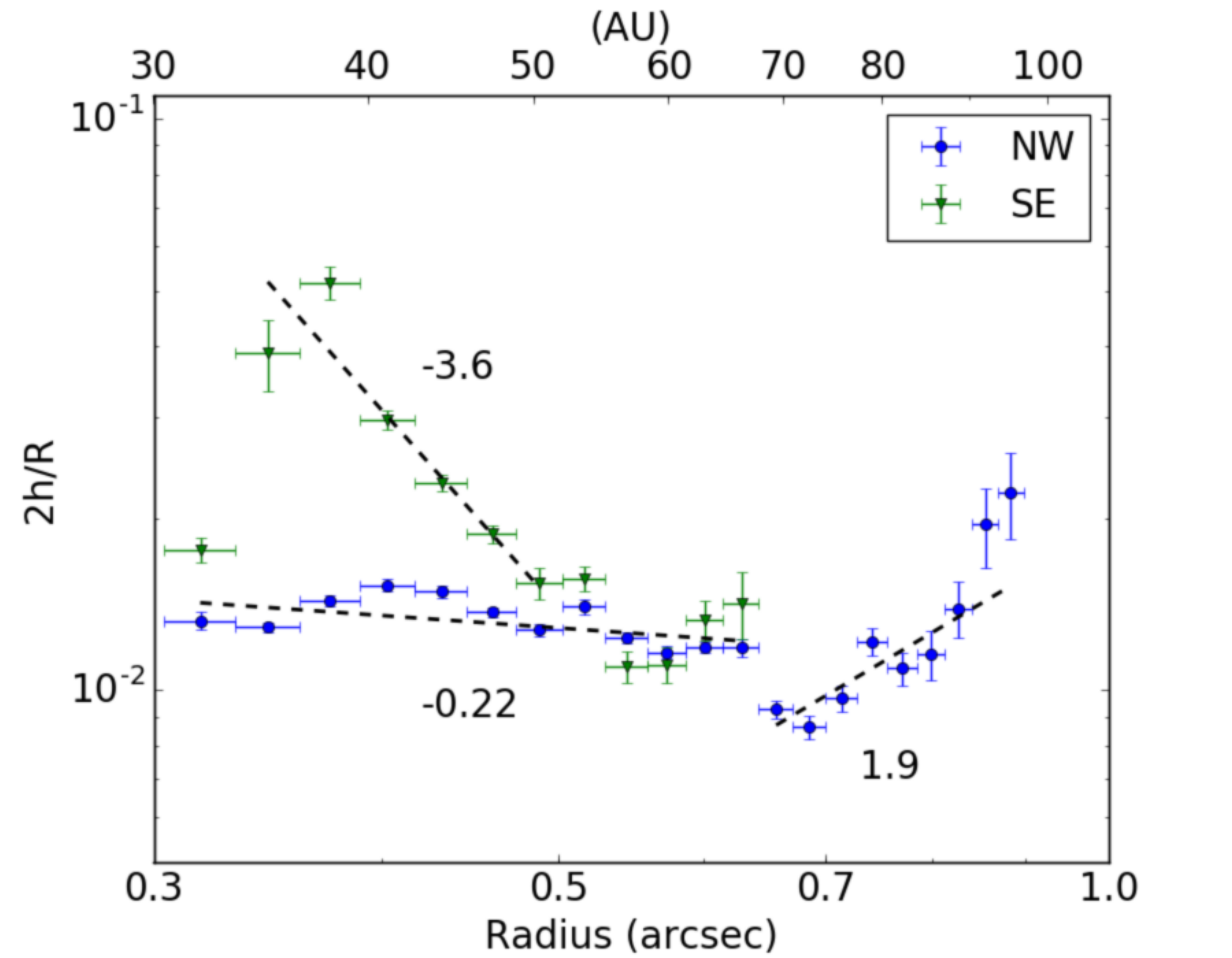}
\caption{Effective vertical FWHM of the disk as a function of stellocentric radius. Green points are the SE extension and blue points are the NW extension. The exponent for each respective power law fit is displayed near the fitted line. An enhancement of the SE extension's scale height relative to the NW extension can be seen inside 50 AU (or $\sim$0\farcs5). Outside that point, it returns to a similar power law with distance.  Unlike the the SE extension, the NW extension is still detected beyond 0\farcs7 and appears to transition to a positive slope.}
\label{fig:sclhgt}
\end{figure}

\section{Morphology}
\label{morph}

In order to measure midplane variations of the disk, we fit a functional profile to the disk emission. It can be seen in Fig. \ref{fig:psf} that all of the PSF subtraction methods show the disk in total intensity. It appears near an inclination of 90$^{\circ}$ and centered on the star. We have added a green reference line passing through both the north and south extensions of the disk. We rotate the PSF-subtracted image by 75$^{\circ}$ clockwise to orient the disk horizontally and to measure the disk emission along the spine relative to the green line at a PA of 165$\degr$. A Cauchy function (Eq. \ref{cauchy}) was fit to the surface brightness ($I$), with a brightness offset ($I_{o}$), and constant ($C$). This technique and function have been used before on edge-on disks such as AU Mic \citep{JG07}. The function was fit along each vertical slice of the disk (about 30 pixels wide) in the $x$ direction, perpendicular to the disk axis, to measure the location of central spine of emission ($x_{o}$) and its FWHM ($\sim$$2h$).

\begin{figure*}[t]

\includegraphics[trim= 0mm 0mm 1.4cm 0mm,height=0.3\textheight,clip]{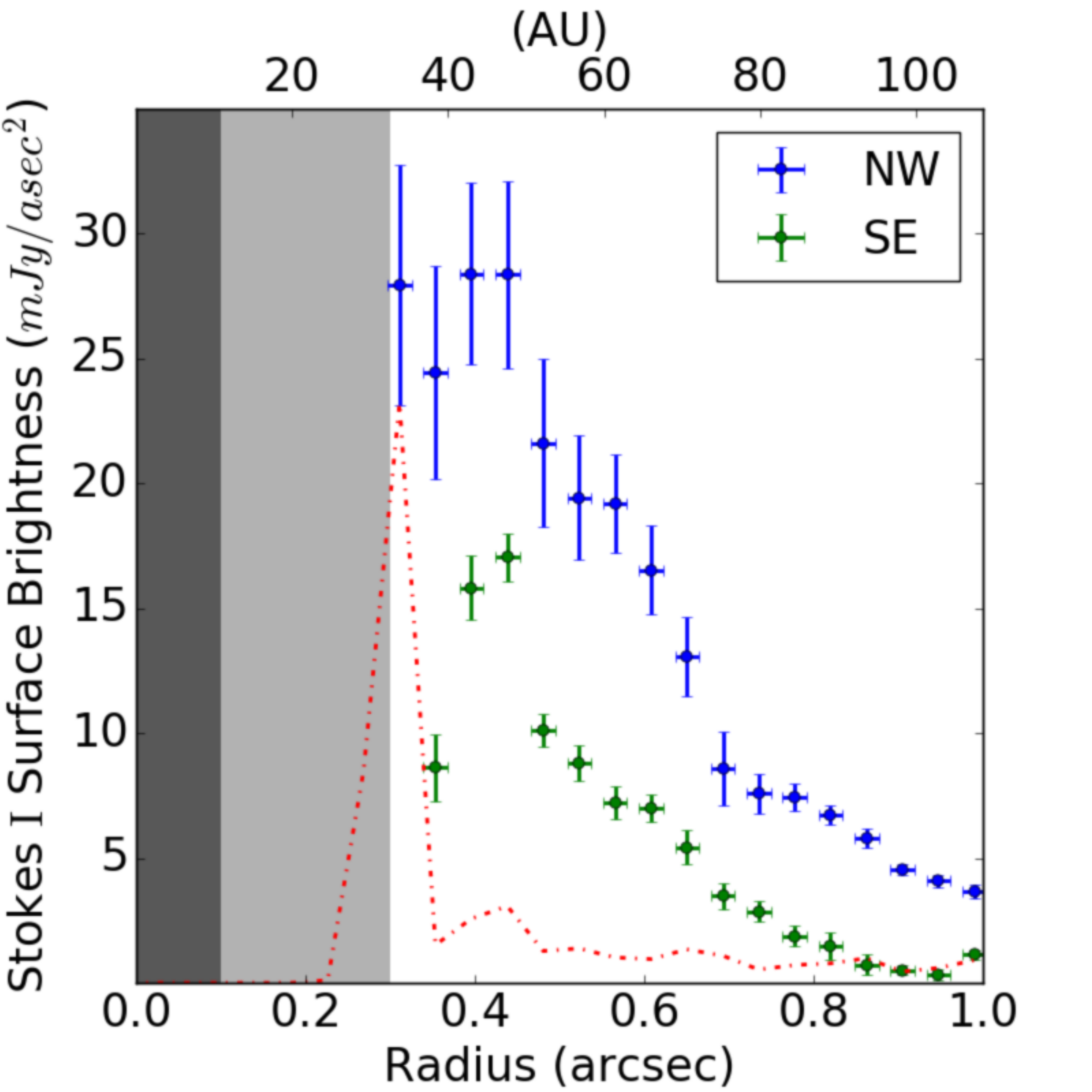}
\includegraphics[trim= 1.6cm 0mm 0mm 0mm,height=0.3\textheight,clip]{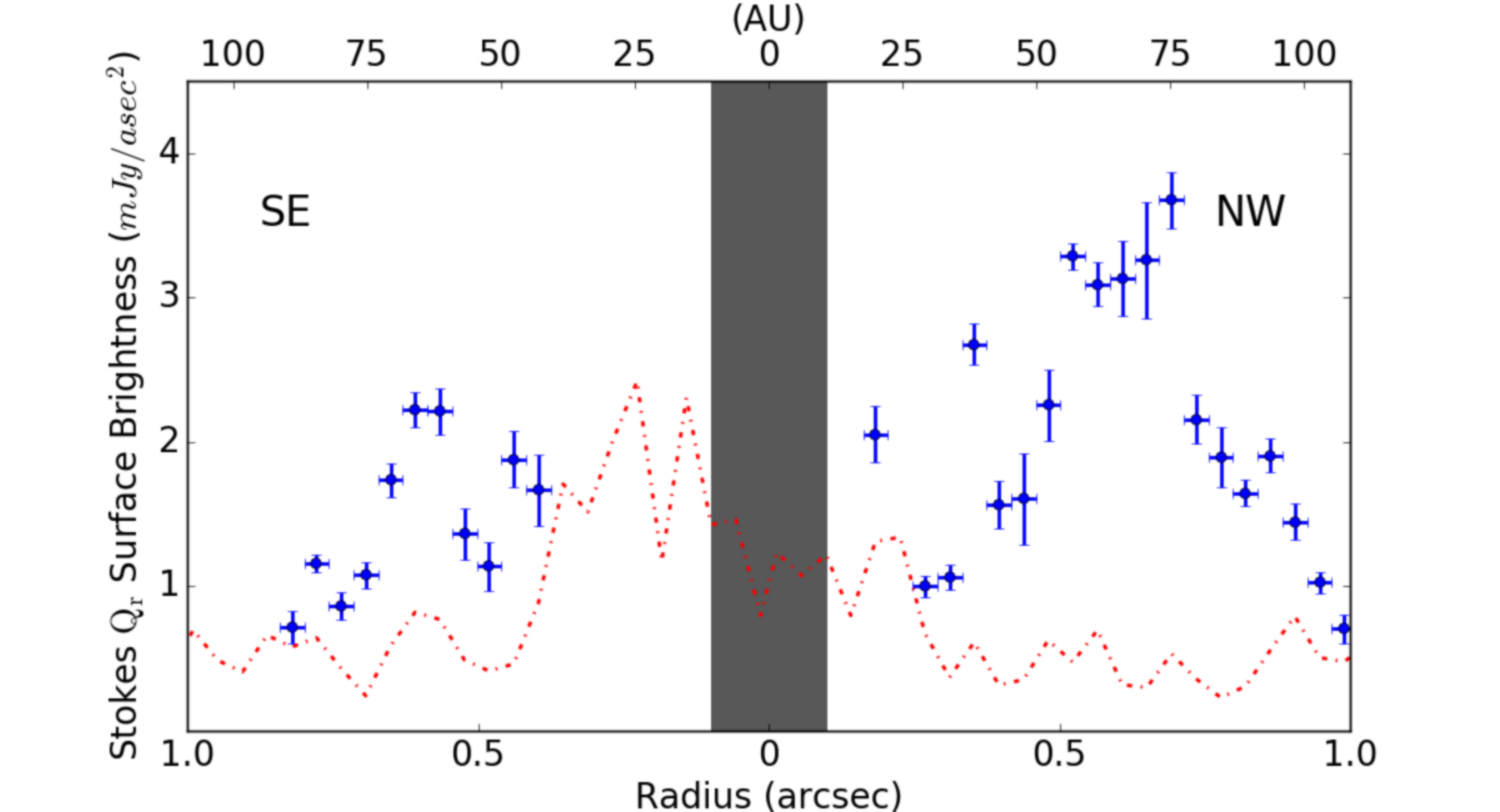}

\caption{$H$-band surface brightness profiles for the combined spectral mode data in total intensity (left panel) and in polarized intensity (right panel). The blue and green dots denote the respective surface brightnesses of 7 pixel wide apertures with standard deviation error bars binned over 5 pixels. The horizontal error bars show the extent of the binned regions. The dotted red line denotes a noise floor. For total intensity, that is mean plus a 1-$\sigma$ standard deviation in regions of the data without a disk (45\degr\ away form the disk midplane). In polarized intensity, it is same except using the $U_r$ image cospatial with the $Q_r$ data.  The dark grey region demarks the area under the coronagraph. The light grey area shows the region inside of the dotted green circle in Fig. \ref{fig:psf} where artifacts from PSF-subtraction are apparent. }

\label{fig:radial}
\end{figure*}

\begin{equation}
\label{cauchy}
I = \frac{C*h}{\pi*(h^{2}+(y-y_{o})^2))}+I_{o}.
\end{equation}

\noindent Fig. \ref{fig:offset} shows the disk mid-plane measurements deviating from $y_0=0$, indicating disk structure from inclination, warping, or both. The $\sim$18 mas offset is significant compared to the upper limit of 1.4 mas astrometric precision. On the SE extension there is a localized offset at 40 AU. If the disk were a symmetric ring inclined close to edge-on, we should see an arc in the disk from one extension to the other (e.g \cite{JM14}; Fig. 5), while if it were perfectly edge-on we should see a flat zero offset for the entire length of the disk through the star's position \citep{PK95}. The deviation in emission along the spine was not significant enough to measure an arc in the disk. Examination leads us to conclude that the disk position angle is 165\degr\ measured to the major disk axis east of north.  If the general offset on the NW extension of the disk is the result of an inclined disk relative to our line of sight, then the offset of the spine of disk emission relative to a line centered through the star would translate to $\sim$1\fdg3--1\fdg7 (assuming a disk ring radius of 70-90 AU), making the disk inclination $\sim$88\degr\ instead of 90\degr. Since inclination and disk position angle can be covariant given these assumptions, they represent general estimates rather than rigorously modeled parameters. The lateral asymmetry again makes it difficult to distinguish between a warp, offset, and/or inclination. 

Furthermore, we examine the projected scale height distribution as a function of separation, in Fig.~\ref{fig:sclhgt}. A scale height enhancement can be seen at the same location as the localized offset in the SE extension inside 50 AU, indicating there is some structure to the disk.  The scale height is measured from the peak emission and independent of any offset.  If the disk is inclined then scale height in this case is rather a projection of emission on the front and back side of the disk. It can be seen that the two sides have different slopes interior to about 0\farcs5 and then have a common slope up until 0\farcs65 where the emission on the SE is noise-dominated, but the NW extension appears to change to a positive slope in scale height, suggesting a transition in the disk emission is occurring around 70 AU, which could be indicative of the location of the disk ansae \citep{JG07}. While the SE disk blob is present in two different PSF subtraction routines, the possibility remains that this observation may be a spurious artifact from emission which is arcing with parallactic rotation close to the noise-dominated region inside 0\farcs3.

\section{Surface Brightness Distribution}
\label{radial}

\begin{figure}
\centering
\includegraphics[width=0.5\textwidth]{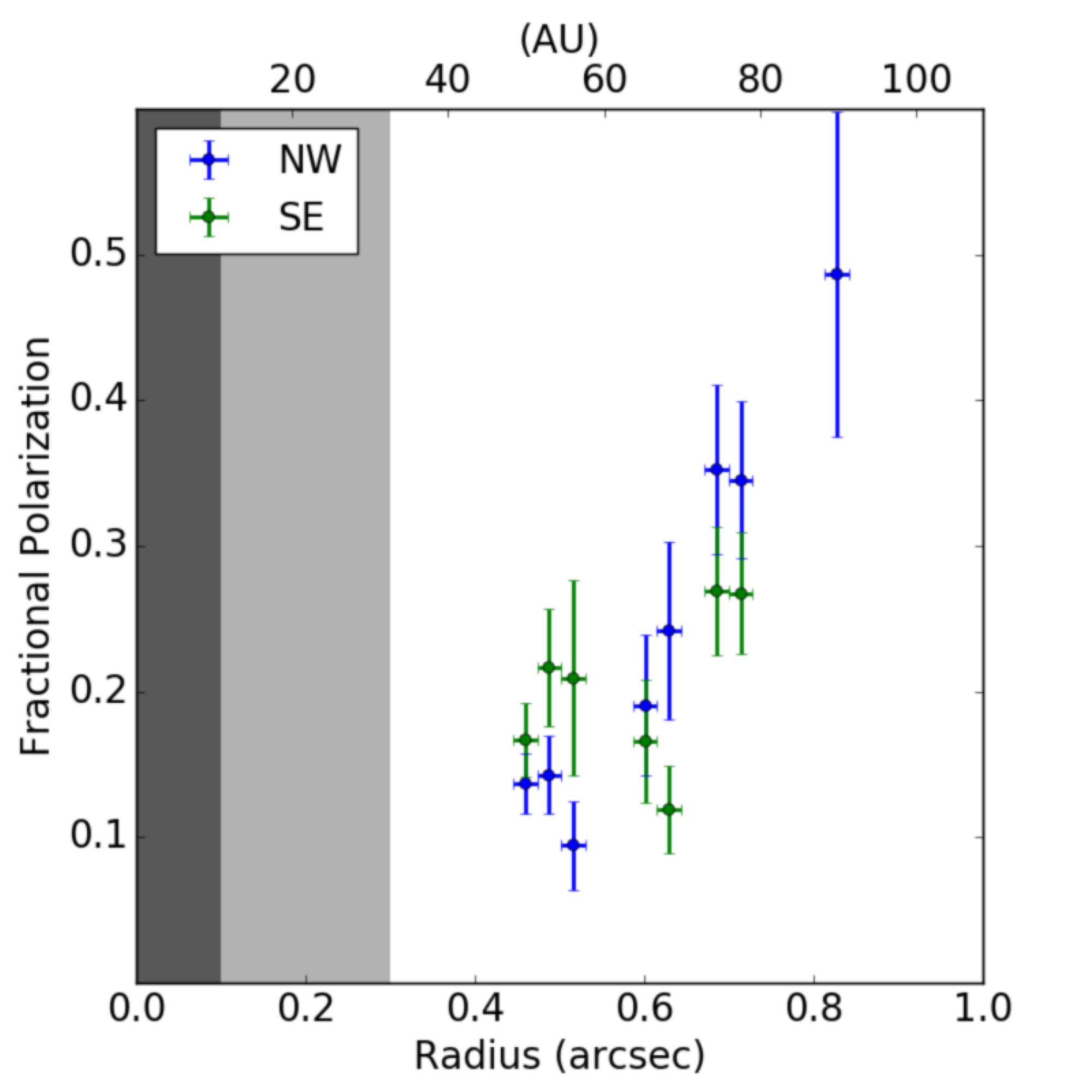}
\caption{Polarization fraction as measured between the spectral mode and polarization mode of GPI. Blue points indicate the NW extension and green points indicate the SE extension of the disk. The polarization fraction trends upward from $\approx$0.1 to $\approx$0.4 in the range of 40 to 80 AU.  Error bars indicate the combined SNR of the spectral mode and polarization mode. Both extensions appear to have similar distribution of fractional polarization with separation from the star given the precision of the current measurements. The polarized intensity dominates the error given a short observing sequence.  Data in regions with total SNR$<$3 are excluded to illustrate where we can confidently measure a fractional polarization.  The dark grey region is the region covered by the coronograph and the light gray region is an area dominated by PSF subtraction artifacts (See Fig. \ref{fig:psf}).}
\label{fig:pfrac}
\end{figure}

In order to determine the brightness of the disk, we again use the masked PSF subtracted images as it results in the least self subtraction of the disk. We rotate the PSF-subtracted image from the disk-masked interpolation method, by 75$^{\circ}$ to orient the disk horizontally within the image to measure the radial surface brightness of the disk (Fig. \ref{fig:radial}).  Using rectangular apertures seven pixels wide in the $y$ (vertical) direction (which is approximately twice the FWHM of a GPI PSF), we measure the surface brightness as a function of distance along the spine of the disk and the standard deviation in each aperture.  The data were binned by averaging every five pixels in the $x$ (horizontal) direction with the errors added in quadrature.  The noise floor of each image is independently estimated by performing the same operation at a PA 45\degr\ away from the disk. Points which are above the red line indicate that the signal in the disk is significant. We find an asymmetry between the SE and the NW side of the disk in total intensity, where the peak intensity on the NW side of the disk is a ratio of 2:1 brighter than the SE side (left panel of Fig. \ref{fig:radial}). The polarized intensity is also about a ratio of 2:1 brighter on the NW side (right panel of Fig. \ref{fig:radial}). Compared to other debris disk, it is one of the most extreme cases of brightness asymmetry as measured at projected separations interior of the inferred ring radius (See \S \ref{morph} \& \ref{sed}).

Overall, the total intensity on the NW side has a smooth decline with radius. The SE side however appears to have a resolved peak near 40 AU, the same location as the scale height enhancement. The NW side similarly appears to flatten around 45--50 AU before being dominated by noise at the inner working angle.  In the polarized intensity, the surface brightness has a pronounced peak stretching from 50 to 75 AU. Different behaviors are expected in the profiles of total intensity and polarized intensity in the context of a ring made of predominantly forward-scattering dust grains. The total intensity along an edge-on disk will be continuously declining with projected separation, with a sharp drop off outside the disk ansae. In contrast, the polarized light may peak in intensity towards increasing scattering angle from the disk. However, this depends on the phase function and the surface density distribution with radius as these two quantities are covariant in total intensity.

With combined total intensity and polarized intensity, it is possible to measure the fractional polarization as a function of separation from the star.  In Fig \ref{fig:pfrac}, it can be seen that, despite the surface brightness asymmetry, the two extensions of the disk follow roughly the same trend upwards to 30\% polarization at 70 AU.  Data are excluded if the combined SNR is within 3$\sigma$ of zero to show only robust detections of the fractional polarization. This largely affects the regions outside the main peaks of polarized intensity from around 50-75 AU, as the noise is dominated by the lower SNR of the polarized intensity detection. A rise in polarization fraction is likely due to a rise towards peak scattering angle near the ansae from an annular disk with an inner gap \citep{JG07}. However, the SNR of our images is insufficient to assess whether a plateau in polarization fraction is achieved within GPI's field-of-view. 

\section{Spectral Energy Distribution}
\label{sed}

In order to provide context for the GPI observations, we fit an SED model to archival photometry of HD~111520 (Fig. \ref{fig:sed}).  Photometry included the optical Tycho-2 survey \citep{EH00} and infrared surveys from 2MASS \citep{CR03} and \textit{Spitzer} \citep{CC14}.  Public archival \textit{Herschel} PACS \citep{pog10} observations (Obs. ID 1342227022-23$;$ PI D. Padgett) and ALMA Cycle 1 Band 6 (1.3~mm) continuum observations (Proj. ID 2012.1.00688.S$;$ PI J. Carpenter), were measured with aperture photometry to better constrain the cold component of the SED (Table \ref{tab:phot}). \textit{Herschel} PACS data was reduced with standard HIPE pipeline \citep{ott10} and was measured with 12 and 22\arcsec\ circular apertures for 70 and 160 $\mu$m with aperture flux correction factors of 0.8 and 0.82, respectively. ALMA continuum maps were retrieved from the ALMA Science Archive and was measured with an 2\farcs5 aperture.  The RMS error was estimated from random apertures of the same size placed in the FOV. Images of the emission associated with HD~111520 can be seen in Fig. \ref{fig:photstamps}.  Whereas, the data are consistent with a point source at 70~$\mu$m, there is some extended emission at 160 $\mu$m and therefore our aperture photometry leads to an over estimate of the 160 $\mu$m flux associated with HD~111520. A second point source is detected in the ALMA map at a PA of 329\degr, 11\farcs9 away from the peak emission of HD 111520 with a flux density of 0.5 mJy. That second source may contribute to the extended emission we see at 160 $\mu$m. It may also be from a background object, but given the perturbed nature of the disk, it is conceivable that it is dynamically relevant, if it were found to be comoving at a separation of $\sim$1200 AU.

\begin{figure}
\centering
\includegraphics[width=0.5\textwidth,trim=0mm 0cm 0mm 1cm,clip]{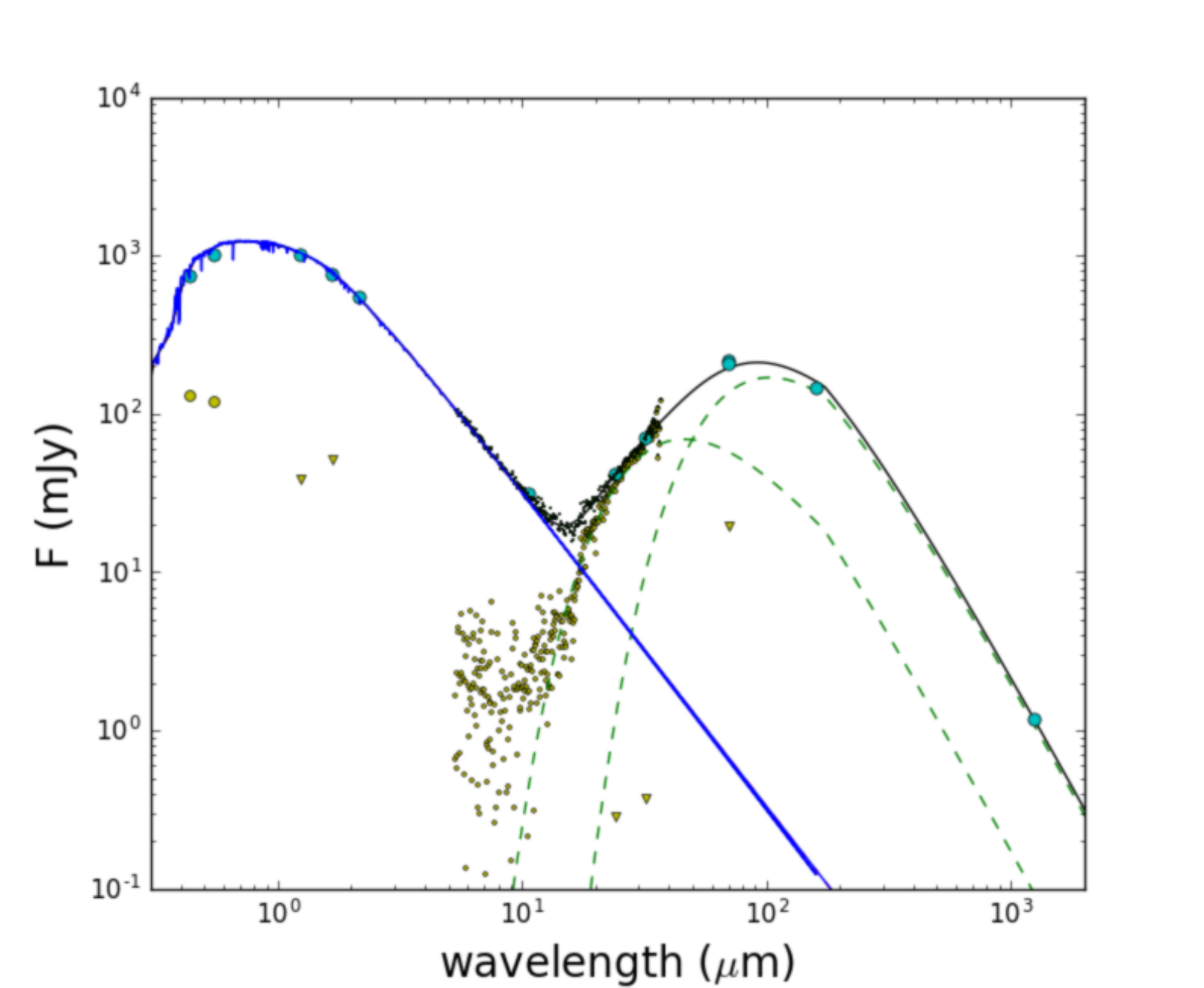}
\caption{Spectral energy distribution for HD 111520.  Archival photometry is in cyan. The black line is the total fit to the data.  The blue line is a stellar Kurucz Model. The two green dashed lines are the modified blackbody dust components. Yellow points denote the residuals of the SED fit with inverted triangles being within measurement uncertainty. {\it Spitzer} IRS spectra are seen as black points.}
\label{fig:sed}
\end{figure}

\begin{figure}
\centering
\includegraphics[width=0.48\textwidth,trim=10mm 0mm 0mm 0mm,clip]{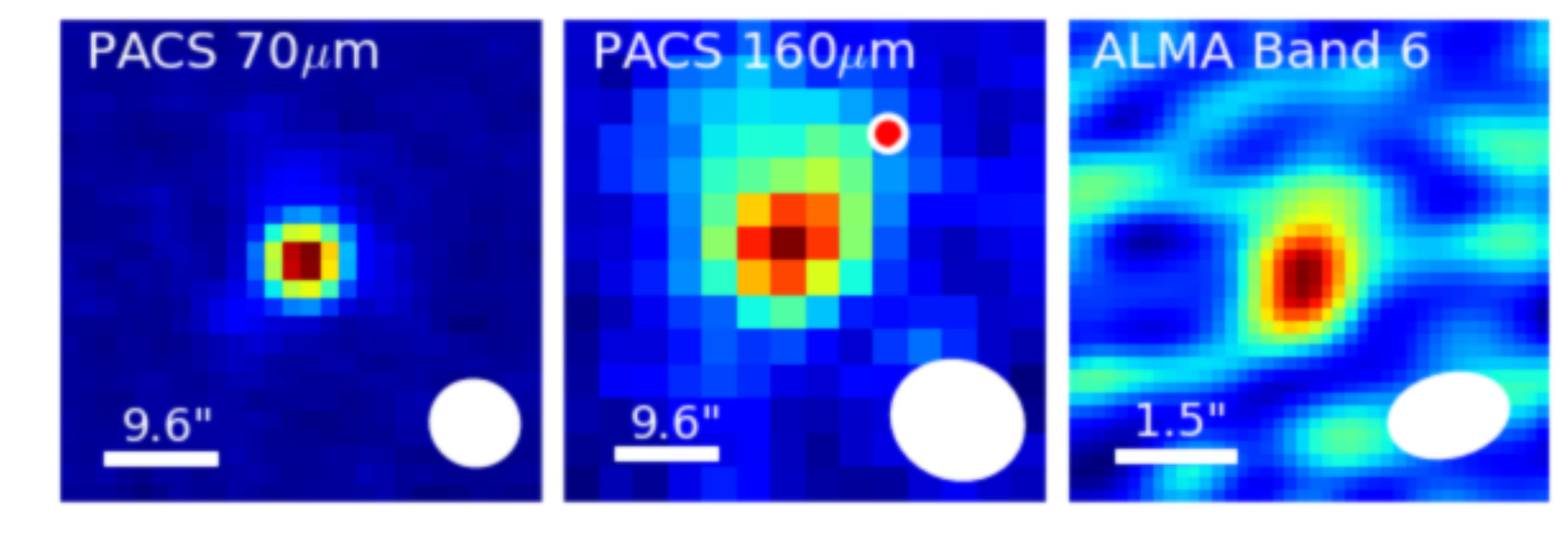}
\caption{Images from \textit{Herschel}/PACS and ALMA showing detections of emission from HD~111520 within their respective wavelengths. The white bars are for image scale and the white ellipses show the respective beam sizes.  At PACS 70$\mu$m and ALMA 1252$\mu$m the emission is seen as a point source while at PACS 160$\mu$m there are hints of extended emission to the N side, possibly stemming from the disk, but possibly due to confusion with other background sources. The location of the second ALMA source is plotted as a red dot in the PACS 160 $\mu$m image. The 160$\mu$m emission seems elongated in a similar direction as the second source, though clearly not all of the flux contamination would be from that source specifically. Note that the ALMA image is shown on a different scale than the PACS observations to best show the emission from HD~111520 and therefore does not reveal the second source.}
\label{fig:photstamps}
\end{figure}

\begin{table}
    \centering
    \caption{Additional photometry from archival observations.}
    \begin{tabular}{ccc}
        Instrument & Effective Wavelength ($\mu$m) & Flux (mJy)  \\
        \hline \\
        \textit{Herschel} PACS & 70 & $205\pm4$ \\
        \textit{Herschel} PACS & 160 & $145\pm6$ \\
        ALMA Band 6 & 1252 & $1.17\pm0.08$ \\
        \hline \\
    \end{tabular}
    \label{tab:phot}
\end{table}

Magnitudes were converted to mJy using the zero points of the respective instruments. \textit{Spitzer} MIPS and \textit{Herschel} PACS have complementary measurements at 70~$\mu$m and are consistent within 1$\sigma$ uncertainties. A Kurucz model was fit to the predominately stellar photometry ($\lambda<$ 10 $\mu$m) in Fig. \ref{fig:sed} with an effective temperature of 6750K. The star subtracted flux densities were then least-squares fit with two modified blackbody SEDs using the photometric uncertainties as weights.  The emission is modified by a power law to model the inefficient emission from grains much smaller than the observed wavelength \citep{MW08}. The modified slope parameters include a knee at 173 $\mu$m with a $\beta$ index of 0.8, but given the lack of photometric coverage near the knee, both parameters remain uncertain. Two components are necessary in order to provide a good fit to all of the data at $\lambda>$10$\mu$m. The temperatures of the warm and cold components in the SED are measured to be 111$\pm$2 K and 49$\pm$2 K, respectively. Uncertainties were determined using the diagonal of the covariance matrix and therefore don't necessarily represent systemic biases such as non-blackbody grains. 

This new SED fit is unique compared to previous SED fitting in that it includes the far-IR observations from \textit{Herschel}, which tightly constrain the temperature of the cold dust component.  Given a stellar luminosity 2.9 $\rm L_{\astrosun}$ and assuming blackbody temperatures for the dust, we find implied disk radii of 11 and 54~AU, respectively, from simple scaling relations \citep{MW08}.  Given that a 11~AU disk component would be completely under the coronograph or dominated by noise, we are mostly resolving emission stemming from the cold component disk.  If the polarized intensity and scale height trends are indicating that the disk radius is near 70 AU and if the scattered light is tracing the population of larger grains from thermal emission, then the disk radius measurements are reasonably consistent.  Since small dust grains are not perfect blackbodies it is not surprising that the actual resolved scattered light radius is larger than the inferred disk radius from the SED fitting \citep{MB13}. The $R_{disk}/R_{BB}$ ratio has been found to scale with luminosity due to radiation pressure more effectively blowing out the smaller grains which have non-blackbody behavior \citep{FM13}.  Applying these relations to this star we would expect the $R_{disk}$ to be 2-3 times that measured by the SED, which is about 108-162 AU or right at the edge of the GPI FOV.

\section{Discussion}

The discovery HST optical images of the HD 111520 disk revealed a nearly perfectly edge-on disk with a strong 5:1 brightness asymmetry \citep{DP15}. Our new $H$-band GPI observations reveal that this asymmetry extends well within the inner working angle of HST, with a 2:1 asymmetry from 0\farcs3 to 1\farcs0. A possible localized brightness enhancement in total intensity at 40 AU is seen on the SE side with two PSF subtraction methods. Explanations for this brightness asymmetry could include a localized variation in dust properties, optical depth effects, or strong density perturbations.

Variations in the dust grains scattering efficiency could cause a variation in brightness if perhaps there were two distinct grain populations on either side of the disk. However, the symmetry of the polarization fraction curve between the two extensions suggests that the dust properties are similar on both ansae. Another possibility is that the dust grains themselves might be at slightly different stellocentric distances resulting from an eccentric disk, possibly induced by a perturbing planet \citep{MW99}. A small brightness asymmetry in thermal emission would then result from the pericenter glow with the brighter side being closer. A similar effect would be observed in scattered light, as shown in potential models of HD 106906 \citep{PK15}. The peak polarized emission on the NW is slightly farther out than the SE side in polarized intensity, suggesting some eccentricity even if we cannot resolve the ansae explicitly. If the disk were eccentric, however, it would cause a brightening on the SE extension rather than the NW extension. Therefore the observed brightness asymmetry cannot be ascribed to localized differences in dust grain properties or disk eccentricity.

Another possibility to consider is that we may just be seeing optical depth effects in the scattered dust. It might be the case that the dust in the outer disk is not asymmetrical, but rather appears that way through disk shadowing. If the inner disk (hotter component) were asymmetrical in scale height, was misaligned relative to the outer disk, or had a locally enhanced density, it could be preferentially shadowing the SE part of the disk. This could occur without needing to invoke a density asymmetry in the outer disk, similar to what is seen in denser protoplanetary disks \citep{CD01,JW08}.  The fractional luminosity of the excess emission ($L_{IR}/L_{*} \approx 10^{-3}$), however, suggests that the scattered light is optically thin and inconsistent with this idea.  Some observations suggest debris disks can still be optically thick in the near-IR such as with HR 4796A \citep{MP15}. In such cases the vertical scale height and width may be narrow enough that a low mass disk could cause shadowing.  However, this would be a transient phenomenon as it would tend to diffuse dynamically into a more diffuse ring.  It may be possible to monitor changes in the inner disk from near-IR variability in concert with scattered light observations to test for transient disk morphology.

If there are density perturbations in the disk such as azimuthal gaps or spirals, when projected at an inclination of 90$^{\circ}$, it would cause a similar brightness variation to what is observed. This would be hard to determine conclusively given our limited viewing angle on the system. HD~111520 itself is an extremely wide binary at a separation of $\sim$159$\arcsec$ (or $\sim$17,000~AU) at a PA of 78\degr\ identified through common proper motion \citep{BM13}. Spiral features induced by a binary star are unlikely, since the co-orbital timescale would be much larger than the orbital timescale of the disk. Smaller mass pertubers have also been searched for with NICI which did not find any low mass companions within 0.5--5$\arcsec$ \citep{MJ13}. It may also be that there is an increased density on the NW side from a recent large collision diffusing small grains, as is seen in $\beta$ Pic for instance \citep{WD14}. Although the sub-mm flux in that case traces the larger grains of the dust whereas we see light being scattered from smaller grains with GPI.  

A few other such systems have been found with similar brightness asymmetries. For example, HD~15115, was discovered to be asymmetric by HST \citep{pk07}.  Using forward modelling of the disk with NICI data, \cite{JM14} were able to show the disk morphology is in fact ring-like at a radius of 90 AU with the east-west asymmetry possibly stemming from either a local over-/under-density or variation in grain properties. It is also thought that an ISM interaction or recent collision of bodies could have occurred and changed the density or size distribution of grains. Another example is HD~106906, which was also shown to be asymmetric in HST data. Images from GPI \citep{PK15} and SPHERE \citep{AL15} show a brightness asymmetry from a near edge-on disk. The variation in brightness is on the order of $\sim$20\% for the total intensity and polarized intensity. A disk which is eccentric, offset, or both, could explain these levels of brightness asymmetry. Since HD~106906 also has a wide orbit planetary companion, it is possible that the dynamical activity between the disk and planet causes this asymmetry. HD~111520 on the other hand has a strong asymmetry throughout the disk (from 5:1 to 2:1), which proves much harder for similar arguments to explain surface brightness variations of that magnitude.  In comparison to the other examples, HD~111520 is the most extreme ``needle-like" disk yet observed.

Given the current data set, it remains impossible to conclusively determine a cause until a more complete picture can be formed through continued monitoring of the system.  What we can determine is that the brightness asymmetry is strong, by a factor of a few, relative to other ``needle''-like debris disks, which are on order of tens of percent. Furthermore it persists from HST observations down to GPI's FOV.  The clump on the SE side will also have to be confirmed and characterized to know if it is relevant to the disk structure. Since the disk has a consistent polarization fraction with distance on both sides, a likely scenario is a large disruption event from a stellar fly-by or planetary perturbations altered the disk density and therefore surface brightness, rather than dust grain inhomogeneities. Through more data of peculiar systems, such as HD 111520, we can determine the true nature and evolution of exo-solar systems. \\

\acknowledgments

Facilities: \facility{Gemini-South}.

Based on observations obtained at the Gemini Observatory, which is operated by the Association of Universities for Research in Astronomy, Inc., under a cooperative agreement with the National Science Foundation (NSF) on behalf of the Gemini partnership: the NSF (United States), the National Research Council (Canada), CONICYT (Chile), the Australian Research Council (Australia), Minist\'{e}rio da Ci\^{e}ncia, Tecnologia e Inova\c{c}\~{a}o (Brazil) and Ministerio de Ciencia, Tecnolog\'{i}a e Inn

This paper makes use of the following ALMA data: ADS/JAO.ALMA \#2012.1.00688.S. ALMA is a partnership of ESO (representing its member states), NSF (USA) and NINS (Japan), together with NRC (Canada) and NSC and ASIAA (Taiwan) and KASI (Republic of Korea), in cooperation with the Republic of Chile. The Joint ALMA Observatory is operated by ESO, AUI/NRAO and NAOJ. The National Radio Astronomy Observatory is a facility of the National Science Foundation operated under cooperative agreement by Associated Universities, Inc.

This research has made use of the SIMBAD database, operated at CDS, Strasbourg, France. 

Z.H.D. and B.C.M. acknowledge a Discovery Grant and Accelerator Supplement from the Natural Science and Engineering Research Council of Canada. 

Supported by NSF grants 
AST-0909188, AST-1313718 (J.R.G., J.J.W., P.G.K.),
AST-141378 (G.D., M.F.), and
AST-1411868 (K.F., J.L.P., A.R., K.W.D.). 

Supported by NASA grants:
NNX15AD95G/NEXSS, NNX14AJ80G and
NNX11AD21G (J.R.G., J.J.W., P.G.K.).

Portions of this work were performed under the auspices of the U.S. Department of Energy by Lawrence Livermore National Laboratory under Contract DE-AC52-07NA27344 (S.M.A.).

\bibliographystyle{apj}
\bibliography{bibtex}

\begin{thebibliography}{}
\expandafter\ifx\csname natexlab\endcsname\relax\def\natexlab#1{#1}\fi

\bibitem[{{Booth} {et~al.}(2013){Booth}, {Kennedy}, {Sibthorpe}, {Matthews},
  {Wyatt}, {Duch{\^e}ne}, {Kavelaars}, {Rodriguez}, {Greaves}, {Koning},
  {Vican}, {Rieke}, {Su}, {Moro-Mart{\'{\i}}n}, \& {Kalas}}]{MB13}
{Booth}, M., {Kennedy}, G., {Sibthorpe}, B., {et~al.} 2013, \mnras, 428, 1263

\bibitem[{{Canovas} {et~al.}(2015){Canovas}, {M{\'e}nard}, {de Boer}, {Pinte},
  {Avenhaus}, \& {Schreiber}}]{HC15}
{Canovas}, H., {M{\'e}nard}, F., {de Boer}, J., {et~al.} 2015, \aap, 582, L7

\bibitem[{{Chen} {et~al.}(2011){Chen}, {Mamajek}, {Bitner}, {Pecaut}, {Su}, \&
  {Weinberger}}]{chen11}
{Chen}, C.~H., {Mamajek}, E.~E., {Bitner}, M.~A., {et~al.} 2011, \apj, 738, 122

\bibitem[{{Chen} {et~al.}(2014){Chen}, {Mittal}, {Kuchner}, {Forrest}, {Lisse},
  {Manoj}, {Sargent}, \& {Watson}}]{CC14}
{Chen}, C.~H., {Mittal}, T., {Kuchner}, M., {et~al.} 2014, \apjs, 211, 25

\bibitem[{{Cutri} {et~al.}(2003){Cutri}, {Skrutskie}, {van Dyk}, {Beichman},
  {Carpenter}, {Chester}, {Cambresy}, {Evans}, {Fowler}, {Gizis}, {Howard},
  {Huchra}, {Jarrett}, {Kopan}, {Kirkpatrick}, {Light}, {Marsh}, {McCallon},
  {Schneider}, {Stiening}, {Sykes}, {Weinberg}, {Wheaton}, {Wheelock}, \&
  {Zacarias}}]{CR03}
{Cutri}, R.~M., {Skrutskie}, M.~F., {van Dyk}, S., {et~al.} 2003, VizieR Online
  Data Catalog, 2246, 0

\bibitem[{{de Zeeuw} {et~al.}(1999){de Zeeuw}, {Hoogerwerf}, {de Bruijne},
  {Brown}, \& {Blaauw}}]{zee99}
{de Zeeuw}, P.~T., {Hoogerwerf}, R., {de Bruijne}, J.~H.~J., {Brown}, A.~G.~A.,
  \& {Blaauw}, A. 1999, \aj, 117, 354

\bibitem[{{Dent} {et~al.}(2014){Dent}, {Wyatt}, {Roberge}, {Augereau},
  {Casassus}, {Corder}, {Greaves}, {de Gregorio-Monsalvo}, {Hales}, {Jackson},
  {Hughes}, {Lagrange}, {Matthews}, \& {Wilner}}]{WD14}
{Dent}, W.~R.~F., {Wyatt}, M.~C., {Roberge}, A., {et~al.} 2014, Science, 343,
  1490

\bibitem[{{Draper} {et~al.}(2014){Draper}, {Marois}, {Wolff}, {Perrin},
  {Ingraham}, {Ruffio}, {Rantakyro}, {Hartung}, \& {Goodsell}}]{ZD14}
{Draper}, Z.~H., {Marois}, C., {Wolff}, S., {et~al.} 2014, in Society of
  Photo-Optical Instrumentation Engineers (SPIE) Conference Series, Vol. 9147,
  Society of Photo-Optical Instrumentation Engineers (SPIE) Conference Series,
  4

\bibitem[{{Dullemond} {et~al.}(2001){Dullemond}, {Dominik}, \& {Natta}}]{CD01}
{Dullemond}, C.~P., {Dominik}, C., \& {Natta}, A. 2001, \apj, 560, 957

\bibitem[{{Ertel} {et~al.}(2012){Ertel}, {Wolf}, \& {Rodmann}}]{SE12}
{Ertel}, S., {Wolf}, S., \& {Rodmann}, J. 2012, \aap, 544, A61

\bibitem[{{Graham} {et~al.}(2007){Graham}, {Kalas}, \& {Matthews}}]{JG07}
{Graham}, J.~R., {Kalas}, P.~G., \& {Matthews}, B.~C. 2007, \apj, 654, 595

\bibitem[{{H{\o}g} {et~al.}(2000){H{\o}g}, {Fabricius}, {Makarov}, {Urban},
  {Corbin}, {Wycoff}, {Bastian}, {Schwekendiek}, \& {Wicenec}}]{EH00}
{H{\o}g}, E., {Fabricius}, C., {Makarov}, V.~V., {et~al.} 2000, \aap, 355, L27

\bibitem[{{Houk}(1978)}]{mss78}
{Houk}, N. 1978, {Michigan catalogue of two-dimensional spectral types for the
  HD stars}

\bibitem[{{Hung} {et~al.}(2015){Hung}, {Duch{\^e}ne}, {Arriaga}, {Fitzgerald},
  {Maire}, {Marois}, {Millar-Blanchaer}, {Bruzzone}, {Rajan}, {Pueyo}, {Kalas},
  {De Rosa}, {Graham}, {Konopacky}, {Wolff}, {Ammons}, {Chen}, {Chilcote},
  {Draper}, {Esposito}, {Gerard}, {Goodsell}, {Greenbaum}, {Hibon}, {Hinkley},
  {Macintosh}, {Marchis}, {Metchev}, {Nielsen}, {Oppenheimer}, {Patience},
  {Perrin}, {Rantakyr{\"o}}, {Sivaramakrishnan}, {Wang}, {Ward-Duong}, \&
  {Wiktorowicz}}]{LH15}
{Hung}, L.-W., {Duch{\^e}ne}, G., {Arriaga}, P., {et~al.} 2015, ArXiv e-prints,
  arXiv:1511.06767

\bibitem[{{Ingraham} {et~al.}(2014{\natexlab{a}}){Ingraham}, {Ruffio},
  {Perrin}, {Wolff}, {Draper}, {Maire}, {Marchis}, \& {Fesquet}}]{PI14b}
{Ingraham}, P., {Ruffio}, J.-B., {Perrin}, M.~D., {et~al.} 2014{\natexlab{a}},
  in \procspie, Vol. 9147, Ground-based and Airborne Instrumentation for
  Astronomy V, 91477K

\bibitem[{{Ingraham} {et~al.}(2014{\natexlab{b}}){Ingraham}, {Perrin},
  {Sadakuni}, {Ruffio}, {Maire}, {Chilcote}, {Larkin}, {Marchis}, {Galicher},
  \& {Weiss}}]{PI14}
{Ingraham}, P., {Perrin}, M.~D., {Sadakuni}, N., {et~al.} 2014{\natexlab{b}},
  in Society of Photo-Optical Instrumentation Engineers (SPIE) Conference
  Series, Vol. 9147, Society of Photo-Optical Instrumentation Engineers (SPIE)
  Conference Series, 7

\bibitem[{{Janson} {et~al.}(2013){Janson}, {Lafreni{\`e}re}, {Jayawardhana},
  {Bonavita}, {Girard}, {Brandeker}, \& {Gizis}}]{MJ13}
{Janson}, M., {Lafreni{\`e}re}, D., {Jayawardhana}, R., {et~al.} 2013, \apj,
  773, 170

\bibitem[{{Kalas} {et~al.}(2007){Kalas}, {Fitzgerald}, \& {Graham}}]{pk07}
{Kalas}, P., {Fitzgerald}, M.~P., \& {Graham}, J.~R. 2007, \apjl, 661, L85

\bibitem[{{Kalas} \& {Jewitt}(1995)}]{PK95}
{Kalas}, P., \& {Jewitt}, D. 1995, \aj, 110, 794

\bibitem[{{Kalas} {et~al.}(2015){Kalas}, {Rajan}, {Wang}, {Millar-Blanchaer},
  {Duchene}, {Chen}, {Fitzgerald}, {Dong}, {Graham}, {Patience}, {Macintosh},
  {Murray-Clay}, {Matthews}, {Rameau}, {Marois}, {Chilcote}, {De Rosa},
  {Doyon}, {Draper}, {Lawler}, {Ammons}, {Arriaga}, {Bulger}, {Cotten},
  {Follette}, {Goodsell}, {Greenbaum}, {Hibon}, {Hinkley}, {Hung}, {Ingraham},
  {Konapacky}, {Lafreniere}, {Larkin}, {Long}, {Maire}, {Marchis}, {Metchev},
  {Morzinski}, {Nielsen}, {Oppenheimer}, {Perrin}, {Pueyo}, {Rantakyr{\"o}},
  {Ruffio}, {Saddlemyer}, {Savransky}, {Schneider}, {Sivaramakrishnan},
  {Soummer}, {Song}, {Thomas}, {Vasisht}, {Ward-Duong}, {Wiktorowicz}, \&
  {Wolff}}]{PK15}
{Kalas}, P.~G., {Rajan}, A., {Wang}, J.~J., {et~al.} 2015, \apj, 814, 32

\bibitem[{{Konopacky} {et~al.}(2014){Konopacky}, {Thomas}, {Macintosh},
  {Dillon}, {Sadakuni}, {Maire}, {Fitzgerald}, {Hinkley}, {Kalas}, {Esposito},
  {Marois}, {Ingraham}, {Marchis}, {Perrin}, {Graham}, {Wang}, {De Rosa},
  {Morzinski}, {Pueyo}, {Chilcote}, {Larkin}, {Fabrycky}, {Goodsell},
  {Oppenheimer}, {Patience}, {Saddlemyer}, \& {Sivaramakrishnan}}]{QK14}
{Konopacky}, Q.~M., {Thomas}, S.~J., {Macintosh}, B.~A., {et~al.} 2014, in
  Society of Photo-Optical Instrumentation Engineers (SPIE) Conference Series,
  Vol. 9147, Society of Photo-Optical Instrumentation Engineers (SPIE)
  Conference Series, 84

\bibitem[{{Lagrange} {et~al.}(2016){Lagrange}, {Langlois}, {Gratton}, {Maire},
  {Milli}, {Olofsson}, {Vigan}, {Bailey}, {Mesa}, {Chauvin}, {Boccaletti},
  {Galicher}, {Girard}, {Bonnefoy}, {Samland}, {Menard}, {Henning},
  {Kenworthy}, {Thalmann}, {Beust}, {Beuzit}, {Brandner}, {Buenzli},
  {Cheetham}, {Janson}, {le Coroller}, {Lannier}, {Mouillet}, {Peretti},
  {Perrot}, {Salter}, {Sissa}, {Wahhaj}, {Abe}, {Desidera}, {Feldt}, {Madec},
  {Perret}, {Petit}, {Rabou}, {Soenke}, \& {Weber}}]{AL15}
{Lagrange}, A.-M., {Langlois}, M., {Gratton}, R., {et~al.} 2016, \aap, 586, L8

\bibitem[{{Macintosh} {et~al.}(2014){Macintosh}, {Graham}, {Ingraham},
  {Konopacky}, {Marois}, {Perrin}, {Poyneer}, {Bauman}, {Barman}, {Burrows},
  {Cardwell}, {Chilcote}, {De Rosa}, {Dillon}, {Doyon}, {Dunn}, {Erikson},
  {Fitzgerald}, {Gavel}, {Goodsell}, {Hartung}, {Hibon}, {Kalas}, {Larkin},
  {Maire}, {Marchis}, {Marley}, {McBride}, {Millar-Blanchaer}, {Morzinski},
  {Norton}, {Oppenheimer}, {Palmer}, {Patience}, {Pueyo}, {Rantakyro},
  {Sadakuni}, {Saddlemyer}, {Savransky}, {Serio}, {Soummer},
  {Sivaramakrishnan}, {Song}, {Thomas}, {Wallace}, {Wiktorowicz}, \&
  {Wolff}}]{BM14}
{Macintosh}, B., {Graham}, J.~R., {Ingraham}, P., {et~al.} 2014, Proceedings of
  the National Academy of Science, 111, 12661

\bibitem[{{Maire} {et~al.}(2014){Maire}, {Ingraham}, {De Rosa}, {Perrin},
  {Rajan}, {Savransky}, {Wang}, {Ruffio}, {Wolff}, {Chilcote}, {Doyon},
  {Graham}, {Greenbaum}, {Konopacky}, {Larkin}, {Macintosh}, {Marois},
  {Millar-Blanchaer}, {Patience}, {Pueyo}, {Sivaramakrishnan}, {Thomas}, \&
  {Weiss}}]{JM14s}
{Maire}, J., {Ingraham}, P.~J., {De Rosa}, R.~J., {et~al.} 2014, in Society of
  Photo-Optical Instrumentation Engineers (SPIE) Conference Series, Vol. 9147,
  Society of Photo-Optical Instrumentation Engineers (SPIE) Conference Series,
  85

\bibitem[{{Marois} {et~al.}(2006){Marois}, {Lafreni{\`e}re}, {Doyon},
  {Macintosh}, \& {Nadeau}}]{CM06}
{Marois}, C., {Lafreni{\`e}re}, D., {Doyon}, R., {Macintosh}, B., \& {Nadeau},
  D. 2006, \apj, 641, 556

\bibitem[{{Mason} {et~al.}(2012){Mason}, {Hartkopf}, \& {Friedman}}]{BM13}
{Mason}, B.~D., {Hartkopf}, W.~I., \& {Friedman}, E.~A. 2012, \aj, 143, 124

\bibitem[{{Matthews} {et~al.}(2014){Matthews}, {Krivov}, {Wyatt}, {Bryden}, \&
  {Eiroa}}]{BM14a}
{Matthews}, B.~C., {Krivov}, A.~V., {Wyatt}, M.~C., {Bryden}, G., \& {Eiroa},
  C. 2014, Protostars and Planets VI, 521

\bibitem[{{Mazoyer} {et~al.}(2014){Mazoyer}, {Boccaletti}, {Augereau},
  {Lagrange}, {Galicher}, \& {Baudoz}}]{JM14}
{Mazoyer}, J., {Boccaletti}, A., {Augereau}, J.-C., {et~al.} 2014, \aap, 569,
  A29

\bibitem[{{Millar-Blanchaer} {et~al.}(2015){Millar-Blanchaer}, {Graham},
  {Pueyo}, {Kalas}, {Dawson}, {Wang}, {Perrin}, {moon}, {Macintosh}, {Ammons},
  {Barman}, {Cardwell}, {Chen}, {Chiang}, {Chilcote}, {Cotten}, {De Rosa},
  {Draper}, {Dunn}, {Duch{\^e}ne}, {Esposito}, {Fitzgerald}, {Follette},
  {Goodsell}, {Greenbaum}, {Hartung}, {Hibon}, {Hinkley}, {Ingraham},
  {Jensen-Clem}, {Konopacky}, {Larkin}, {Long}, {Maire}, {Marchis}, {Marley},
  {Marois}, {Morzinski}, {Nielsen}, {Palmer}, {Oppenheimer}, {Poyneer},
  {Rajan}, {Rantakyr{\"o}}, {Ruffio}, {Sadakuni}, {Saddlemyer}, {Schneider},
  {Sivaramakrishnan}, {Soummer}, {Thomas}, {Vasisht}, {Vega}, {Wallace},
  {Ward-Duong}, {Wiktorowicz}, \& {Wolff}}]{MB15}
{Millar-Blanchaer}, M.~A., {Graham}, J.~R., {Pueyo}, L., {et~al.} 2015, \apj,
  811, 18

\bibitem[{{Mittal} {et~al.}(2015){Mittal}, {Chen}, {Jang-Condell}, {Manoj},
  {Sargent}, {Watson}, \& {Lisse}}]{TM15}
{Mittal}, T., {Chen}, C.~H., {Jang-Condell}, H., {et~al.} 2015, \apj, 798, 87

\bibitem[{{Morales} {et~al.}(2013){Morales}, {Bryden}, {Werner}, \&
  {Stapelfeldt}}]{FM13}
{Morales}, F.~Y., {Bryden}, G., {Werner}, M.~W., \& {Stapelfeldt}, K.~R. 2013,
  \apj, 776, 111

\bibitem[{{Mustill} \& {Wyatt}(2009)}]{AM09}
{Mustill}, A.~J., \& {Wyatt}, M.~C. 2009, \mnras, 399, 1403

\bibitem[{{Ott}(2010)}]{ott10}
{Ott}, S. 2010, in Astronomical Society of the Pacific Conference Series, Vol.
  434, Astronomical Data Analysis Software and Systems XIX, ed. Y.~{Mizumoto},
  K.-I. {Morita}, \& M.~{Ohishi}, 139

\bibitem[{{Padgett} \& {Stapelfeldt}(2015)}]{DP15}
{Padgett}, D., \& {Stapelfeldt}, K. 2015, in IAU Symposium, Vol. 314, Young
  Stars \& Planets Near the Sun, ed. J.~H. {Kastner}, B.~{Stelzer}, \& S.~A.
  {Metchev}, 175--178

\bibitem[{{Pecaut} {et~al.}(2012){Pecaut}, {Mamajek}, \& {Bubar}}]{pec12}
{Pecaut}, M.~J., {Mamajek}, E.~E., \& {Bubar}, E.~J. 2012, \apj, 746, 154

\bibitem[{{Perrin} {et~al.}(2014){Perrin}, {Maire}, {Ingraham}, {Savransky},
  {Millar-Blanchaer}, {Wolff}, {Ruffio}, {Wang}, {Draper}, {Sadakuni},
  {Marois}, {Rajan}, {Fitzgerald}, {Macintosh}, {Graham}, {Doyon}, {Larkin},
  {Chilcote}, {Goodsell}, {Palmer}, {Labrie}, {Beaulieu}, {De Rosa},
  {Greenbaum}, {Hartung}, {Hibon}, {Konopacky}, {Lafreniere}, {Lavigne},
  {Marchis}, {Patience}, {Pueyo}, {Rantakyr{\"o}}, {Soummer},
  {Sivaramakrishnan}, {Thomas}, {Ward-Duong}, \& {Wiktorowicz}}]{per14}
{Perrin}, M.~D., {Maire}, J., {Ingraham}, P., {et~al.} 2014, in Society of
  Photo-Optical Instrumentation Engineers (SPIE) Conference Series, Vol. 9147,
  Society of Photo-Optical Instrumentation Engineers (SPIE) Conference Series,
  3

\bibitem[{{Perrin} {et~al.}(2015){Perrin}, {Duchene}, {Millar-Blanchaer},
  {Fitzgerald}, {Graham}, {Wiktorowicz}, {Kalas}, {Macintosh}, {Bauman},
  {Cardwell}, {Chilcote}, {De Rosa}, {Dillon}, {Doyon}, {Dunn}, {Erikson},
  {Gavel}, {Goodsell}, {Hartung}, {Hibon}, {Ingraham}, {Kerley}, {Konapacky},
  {Larkin}, {Maire}, {Marchis}, {Marois}, {Mittal}, {Morzinski}, {Oppenheimer},
  {Palmer}, {Patience}, {Poyneer}, {Pueyo}, {Rantakyr{\"o}}, {Sadakuni},
  {Saddlemyer}, {Savransky}, {Soummer}, {Sivaramakrishnan}, {Song}, {Thomas},
  {Wallace}, {Wang}, \& {Wolff}}]{MP15}
{Perrin}, M.~D., {Duchene}, G., {Millar-Blanchaer}, M., {et~al.} 2015, \apj,
  799, 182

\bibitem[{{Poglitsch} {et~al.}(2010){Poglitsch}, {Waelkens}, {Geis},
  {Feuchtgruber}, {Vandenbussche}, {Rodriguez}, {Krause}, {Renotte}, {van
  Hoof}, {Saraceno}, {Cepa}, {Kerschbaum}, {Agn{\`e}se}, {Ali}, {Altieri},
  {Andreani}, {Augueres}, {Balog}, {Barl}, {Bauer}, {Belbachir}, {Benedettini},
  {Billot}, {Boulade}, {Bischof}, {Blommaert}, {Callut}, {Cara}, {Cerulli},
  {Cesarsky}, {Contursi}, {Creten}, {De Meester}, {Doublier}, {Doumayrou},
  {Duband}, {Exter}, {Genzel}, {Gillis}, {Gr{\"o}zinger}, {Henning},
  {Herreros}, {Huygen}, {Inguscio}, {Jakob}, {Jamar}, {Jean}, {de Jong},
  {Katterloher}, {Kiss}, {Klaas}, {Lemke}, {Lutz}, {Madden}, {Marquet},
  {Martignac}, {Mazy}, {Merken}, {Montfort}, {Morbidelli}, {M{\"u}ller},
  {Nielbock}, {Okumura}, {Orfei}, {Ottensamer}, {Pezzuto}, {Popesso},
  {Putzeys}, {Regibo}, {Reveret}, {Royer}, {Sauvage}, {Schreiber}, {Stegmaier},
  {Schmitt}, {Schubert}, {Sturm}, {Thiel}, {Tofani}, {Vavrek}, {Wetzstein},
  {Wieprecht}, \& {Wiezorrek}}]{pog10}
{Poglitsch}, A., {Waelkens}, C., {Geis}, N., {et~al.} 2010, \aap, 518, L2

\bibitem[{{Pueyo} {et~al.}(2015){Pueyo}, {Soummer}, {Hoffmann}, {Oppenheimer},
  {Graham}, {Zimmerman}, {Zhai}, {Wallace}, {Vescelus}, {Veicht}, {Vasisht},
  {Truong}, {Sivaramakrishnan}, {Shao}, {Roberts}, {Roberts}, {Rice}, {Parry},
  {Nilsson}, {Lockhart}, {Ligon}, {King}, {Hinkley}, {Hillenbrand}, {Hale},
  {Dekany}, {Crepp}, {Cady}, {Burruss}, {Brenner}, {Beichman}, \&
  {Baranec}}]{LP15}
{Pueyo}, L., {Soummer}, R., {Hoffmann}, J., {et~al.} 2015, \apj, 803, 31

\bibitem[{{Schmid} {et~al.}(2006){Schmid}, {Joos}, \& {Tschan}}]{HS06}
{Schmid}, H.~M., {Joos}, F., \& {Tschan}, D. 2006, \aap, 452, 657

\bibitem[{{Soummer} {et~al.}(2012){Soummer}, {Pueyo}, \& {Larkin}}]{RS12}
{Soummer}, R., {Pueyo}, L., \& {Larkin}, J. 2012, \apjl, 755, L28

\bibitem[{{van Leeuwen}(2007)}]{lee07}
{van Leeuwen}, F. 2007, \aap, 474, 653

\bibitem[{{Wang} {et~al.}(2015){Wang}, {Ruffio}, {De Rosa}, {Aguilar}, {Wolff},
  \& {Pueyo}}]{JW15}
{Wang}, J.~J., {Ruffio}, J.-B., {De Rosa}, R.~J., {et~al.} 2015, {pyKLIP: PSF
  Subtraction for Exoplanets and Disks}, Astrophysics Source Code Library,
  ascl:1506.001

\bibitem[{{Wang} {et~al.}(2014){Wang}, {Rajan}, {Graham}, {Savransky},
  {Ingraham}, {Ward-Duong}, {Patience}, {De Rosa}, {Bulger},
  {Sivaramakrishnan}, {Perrin}, {Thomas}, {Sadakuni}, {Greenbaum}, {Pueyo},
  {Marois}, {Oppenheimer}, {Kalas}, {Cardwell}, {Goodsell}, {Hibon}, \&
  {Rantakyr{\"o}}}]{JW14}
{Wang}, J.~J., {Rajan}, A., {Graham}, J.~R., {et~al.} 2014, in Society of
  Photo-Optical Instrumentation Engineers (SPIE) Conference Series, Vol. 9147,
  Society of Photo-Optical Instrumentation Engineers (SPIE) Conference Series,
  55

\bibitem[{{Wisniewski} {et~al.}(2008){Wisniewski}, {Clampin}, {Grady},
  {Ardila}, {Ford}, {Golimowski}, {Illingworth}, \& {Krist}}]{JW08}
{Wisniewski}, J.~P., {Clampin}, M., {Grady}, C.~A., {et~al.} 2008, \apj, 682,
  548

\bibitem[{{Wolff} {et~al.}(2014){Wolff}, {Perrin}, {Maire}, {Ingraham},
  {Rantakyr{\"o}}, \& {Hibon}}]{SW14}
{Wolff}, S.~G., {Perrin}, M.~D., {Maire}, J., {et~al.} 2014, in Society of
  Photo-Optical Instrumentation Engineers (SPIE) Conference Series, Vol. 9147,
  Society of Photo-Optical Instrumentation Engineers (SPIE) Conference Series,
  7

\bibitem[{{Wyatt}(2008)}]{MW08}
{Wyatt}, M.~C. 2008, \araa, 46, 339

\bibitem[{{Wyatt} {et~al.}(1999){Wyatt}, {Dermott}, {Telesco}, {Fisher},
  {Grogan}, {Holmes}, \& {Pi{\~n}a}}]{MW99}
{Wyatt}, M.~C., {Dermott}, S.~F., {Telesco}, C.~M., {et~al.} 1999, \apj, 527,
  918

\end{thebibliography}

\end{document}